\documentclass{article}
\pdfoutput=1 

\usepackage{jcappub} 

\usepackage{xcolor}
\usepackage{hyperref,url}
\usepackage{hhline, array, multirow, makecell,xfrac,array}
\usepackage{aas_macros}
\newcommand{\lcdm}{$\Lambda$CDM}
\newcommand{\bc}{\mathbf{C}_\ell}
\newcommand{\omb}{\omega_b}
\newcommand{\omc}{\omega_c}
\newcommand{\treio}{\tau_\mathrm{reio}}
\newcommand{\geff}{G_\mathrm{eff}}
\newcommand{\gefft}{\tilde{G}_\mathrm{eff}}
\newcommand{\lgeff}{\log_{10} G_\mathrm{eff}}
\newcommand{\lgefft}{\log_{10} \tilde{G}_\mathrm{eff}}
\newcommand{\lmax}{\ell_\mathrm{max}}
\newcommand{\neff}{N_\mathrm{eff}}
\newcommand{\dneff}{\Delta N_\mathrm{eff}}
\newcommand{\muKam}{\,\mu\mathrm{K\text{-}arcmin}}

\newcommand{\three}{$\mathbf {3c+0f}$}
\newcommand{\two}{$\mathbf {2c+1f}$}
\newcommand{\one}{$\mathbf {1c+2f}$}

\newcolumntype{P}[1]{>{\centering\arraybackslash}p{#1}}
\newcolumntype{M}[1]{>{\centering\arraybackslash}m{#1}}

\title{\boldmath The magnificent ACT of flavor-specific neutrino self-interaction}

\author[a,b]{Anirban Das}
\author[c]{and Subhajit Ghosh}

\affiliation[a]{SLAC National Accelerator Laboratory, 2575 Sand Hill Road, Menlo Park, California 94025, USA}
\affiliation[b]{Center for Theoretical Physics, Department of Physics and Astronomy, Seoul National University, Seoul 08826, South Korea}
\affiliation[c]{Department of Physics, University of Notre Dame,
South Bend, IN 46556, USA}

\emailAdd{anirbandas@snu.ac.kr}
\emailAdd{sghosh5@nd.edu}
\preprint{SLAC-PUB-17708}

\abstract{We revisit the cosmology of neutrino self-interaction and use the latest cosmic microwave background data from the Atacama Cosmology Telescope (ACT) and the Planck experiment to constrain the interaction strength. In both flavor-universal and nonuniversal coupling scenarios, we find that the ACT data prefers strong neutrino self-interaction that delays neutrino free streaming until just before the matter-radiation equality. When combined with the Planck 2018 data, the preference for strong interaction decreases due to the Planck polarization data. For the combined dataset, the flavor-specific interaction still provides a better fit to the CMB data than $\Lambda$CDM. This trend persists even when neutrino mass is taken into account and extra radiation is added. We also study the prospect of constraining such strong interaction by future terrestrial and space telescopes, and find that the upcoming CMB-S4 experiment will improve the upper limit on neutrino self-interaction by about a factor of three.}

\begin{document} 
\maketitle
\flushbottom

\section{Introduction}
\label{sec:intro}
Observational cosmology has made great achievements over the last three decades, especially in the frontier of cosmic microwave background (CMB) observation. Starting from the COBE-FIRAS experiment in the early 1990s, continuing with the WMAP satellite in the 2000s, and finally, the Planck experiment in the 2010s have continued the legacy measurement of the CMB anisotropy power spectra with ever-increasing precision between generations. Just within the past 30 years, the temperature measurement sensitivity of the instruments has improved by about 100 times which has resulted in a $~\sim 0.1\%$ precision on anisotropy measurement. Today, even though Planck has released its final data in 2018, other experiments are continuing to observe the microwave sky and more missions are being designed for the future.

Currently, multiple terrestrial CMB experiments are ongoing, such as, Atacama Cosmology Telescope (ACT)\,\cite{ACT:2020gnv} in Chile, South Pole Telescope-3G (SPT-3G)\,\cite{SPT-3G:2021eoc}, and BICEP3\,\cite{BICEP3:2016pqy} at the South Pole. While Planck has been able to measure the CMB power spectrum with $\mathcal{O}(50\,\muKam)$ level sensitivity and few arcmin angular resolution, its data goes only upto $\lmax=2500$ for temperature and 2000 for polarization spectrum\,\cite{Planck:2018nkj,Planck:2018vyg}. Measuring the polarization power spectra with better precision and at smaller scales is one of the key goals of ongoing and future missions. For example, ACT collaboration has published temperature and polarization power spectra with the angular resolution a few arcmins (corresponding to $\lmax\simeq 4000$), which is a major improvement over the Planck data\,\cite{ACT:2020gnv}. The SPT-3G has measured the spectra at the level of $\lmax=3000$\,\cite{SPT-3G:2021eoc}. In combination with the Planck data, these experiments have helped us improve the measurement of the parameters of the \lcdm{} model of cosmology and constrain its various extensions.

Looking into the future, there are multiple telescopes that are being built or planned. CMB-S4 is a terrestrial experiment that has been approved and is in its final phase of technical designing. One of the key goals of CMB-S4 is to achieve sub-arcmin angular resolution and unprecedented sensitivity to measure the small-scale power spectrum\,\cite{Abazajian:2019eic}. It is likely to use both the Chilean and the South Pole sites to perform ultra-deep and wide-field surveys. A space-based mission CORE has been proposed in the European Space Agency's M5 call for mid-sized missions in 2016\,\cite{CORE:2017oje}. According to their plan, CORE is going to focus on small-scale polarization data. PICO has been proposed with an ambitious goal of surveying the whole CMB sky in a broad range of frequencies (21-799 GHz) with a sensitivity about 100 times better than the Planck experiment\,\cite{NASAPICO:2019thw}. Another space-based mission LiteBIRD is being set to launch later in this decade to measure the B-mode polarization of the CMB\,\cite{LiteBIRD:2020khw}. The BICEP array is being built to measure the B-mode polarization signal with greater precision and will replace the Keck array\,\cite{BICEP3:2016pqy}. More futuristic experiments like CMB-HD, if materialized according to its current plan, would provide a remarkably valuable tool to look for new physics in CMB, thanks to its significantly better angular resolution and improved sensitivity, and help us detect or constrain beyond-\lcdm{} models\,\cite{CMB-HD:2022bsz}.

Many extensions of the \lcdm{} have been considered as alternative models of the Universe. Among them, models with extra relativistic degrees of freedom in the form of neutrinos and/or other light dark sector particles with secret self-interaction are particularly interesting as they often arise in many particle physics models. Moreover, neutrinos hold a guaranteed key to the discovery of a new beyond Standard Model physics with its mass not explained by the current theory.
The terrestrial neutrino oscillation experiments have improved the measurements of the neutrino mixing parameters over the last few decades\,\cite{Esteban:2020cvm}. The existence of neutrino flavor oscillation proves that they have nonzero mass. However, the absolute mass scale of the neutrinos and any knowledge about self-interaction other than the Standard Model weak force remains unknown to us. Future CMB measurements might hold the key to success in this area. 

Neutrino self-interaction has been a topic of active research since the early days of CMB observation\,\cite{Raffelt:1987ah,Atrio-Barandela:1996suw,Hannestad:2000gt}. The existence of free-streaming relativistic species in the early Universe affects the observed CMB anisotropy power spectra by suppressing its amplitude and shifting it toward larger angular scales. This happens due to the large relativistic speed of the neutrinos, the effect of which is transferred to other species via gravity. However, if there is self-interaction among the particles, then it impedes their free-streaming and lessens these effects on the power spectra. This in turn cancels the phase shift and boosts the power spectra at the small angular scale compared to the free streaming scenario. The new effects of neutrino self-interaction have been a topic of great interest over the last several years in various contexts including the so-called Hubble tension\,\cite{Cyr-Racine:2013jua,Dasgupta:2013zpn,Hannestad:2013ana,Archidiacono:2014nda,Das:2017iuj,Kreisch:2019yzn,DeGouvea:2019wpf,Das:2020xke,RoyChoudhury:2020dmd,Brinckmann:2020bcn,Das:2021guu,Venzor:2022hql,Das:2022xsz,Chang:2022aas,Kreisch:2022zxp,Ghosh:2018usj,Loverde:2022wih,Berger:2022cab,Barenboim:2019tux,Corona:2021qxl,RoyChoudhury:2022rva} For more details on self-interacting neutrino, a recent comprehensive review can be found in Ref.\,\cite{Berryman:2022hds}. However, for an alternative interpretation of the Hubble tension, see Ref.~\cite{Rameez:2019wdt,Colgain:2022nlb} for example.

The effects of self-interacting neutrino (SINU) are present at scales that entered the horizon long before the onset of their free streaming which typically happens before recombination in the models of our interest. 
Therefore, very precise measurements of the CMB power spectra at smaller angular scales are crucial. The recent result from the ACT, which extends up to multipole $\ell\simeq4000$ with significantly improved precision than Planck, is a huge progress in this direction.
In this work, we use this latest ACT data to update the constraint on both flavor-universal and nonuniversal self-interaction using massless neutrinos. We also combine ACT with the 2018 Planck data to investigate their combined constraining power. In concordance with earlier works, we find a strongly interacting (SI) and a moderately interacting (MI) mode in the posterior distribution of the coupling strength.
We find that ACT prefers a slightly stronger self-interaction for the SI mode compared to Planck. It also increases the overall significance of the SI mode relative to a Planck-only analysis. These observations could be attributed to a specific feature in the ACT data. The results for the strong coupling mode remain unchanged qualitatively when a nonzero neutrino mass is included with expected changes in a few other cosmological parameters showing the validity of our baseline analysis with massless neutrinos.  We also study an extension of the \lcdm{} model of cosmology by varying the number of extra relativistic degrees of freedom. The extra radiation does not affect the neutrino self-coupling strength in any significant manner except changing some of the background parameters.

Finally, we perform a Fisher forecast analysis of SINU cosmology for a few future CMB experiments, such as CMB-S4 and CORE. We point out the multipole range that contains the most information about SINU, and find that CMB-S4 will improve the constraint on the coupling strength approximately by a factor of three.
This future forecast about neutrino self-interaction for terrestrial and space-based experiments would be useful in planning future experiments. 

The outline of the paper is as follows. In section\,\ref{sec:sinu_primer}, we give a brief introduction to the cosmology of self-interacting neutrino. We show our data analysis and the results in Sec.\,\ref{sec:results} followed by the future prospect analysis in Sec.\,\ref{sec:future}. Finally, we discuss the results in this paper and conclude in Sec.\,\ref{sec:conclusion}.

\section{Cosmology of self-interacting neutrino}
\label{sec:sinu_primer}
Following our previous work in Ref.\,\cite{Das:2020xke}, we consider a phenomenological model of self-interacting neutrino (SINU) where the interaction is mediated by a heavy mediator. This means the mediator mass is greater than the typical momentum transfer in the scattering during all relevant cosmological epochs for CMB. In this case, the interaction term in the Lagrangian can be written as a four-Fermi effective operator,
\begin{equation}
    \mathcal{L}\supset G^{(ijkl)}_\mathrm{eff} \bar{\nu}_i\nu_j\bar{\nu}_k\nu_l\,,
\end{equation}
where $i,j,k,l$ are neutrino flavor indices, and $G^{(ijkl)}_\mathrm{eff}$ is the effective coupling strength for such states. This coupling is proportional to the inverse square of the mediator mass.
In this work, we will consider only flavor-diagonal coupling matrices,
\begin{equation}
    \geff = \mathrm{diag}\{\geff^{(1)}, \geff^{(2)},\ldots \geff^{(N)}\}\,.
\end{equation}
These diagonal coupling strengths will be free parameters. We will also always take the self-interacting neutrinos to be massless. As a result, there is no mixing between different flavors in this scenario. Hence, we are free to define flavor-universal or nonuniversal interactions in this basis without worrying about the basis transformation of the coupling matrix. 

With this effective coupling, neutrino scattering cross section is $\langle\sigma v\rangle \propto (\geff)^2T^2$, and the scattering rate is $\Gamma\propto (\geff)^2T_\nu^5$ with $T_\nu$ as the neutrino bath temperature. Therefore, the covariant neutrino opacity $\dot{\tau}_\nu$ due to self-scattering can be defined as 
\begin{equation}
\label{eq:taudot}
    \dot{\tau}_\nu \equiv -a(\geff)^2T_\nu^5\,.
\end{equation}
Here, $a$ is the scale factor of the Universe.
In the presence of self-coupling, the neutrino perturbation equations for the anisotropic stress and higher multipoles in the Boltzmann hierarchy are augmented with a collision term proportional to $\dot{\tau}_\nu$. This collision term has a damping action on the neutrino propagation before their decoupling and results in a late onset of neutrino free streaming compared to \lcdm. In the absence of self-interaction, the neutrinos travel with a supersonic speed after decoupling, dragging the photon-baryon plasma with it through gravitational interaction. However, in the presence of strong self-interaction, the late onset of free streaming changes this and affects the growth of fluctuations in the photons and baryons. The SINU induced changes in the CMB and matter power spectra are discussed in detail in Ref.\,\cite{Baumann:2015rya,Kreisch:2019yzn,Das:2020xke}. To summarize, SINU causes those perturbations, that entered the horizon before neutrino decoupling, to grow more than in \lcdm{} which enhances the CMB temperature and polarization anisotropy power spectra. Additionally, it causes a phase shift of the oscillations in the power spectra toward smaller angular scales by changing the distance to the last scattering surface from us. 

Both the effects on power amplitude and phase shift are proportional to the fraction of the free streaming radiation energy density which is quantified by $R_\nu$\,\cite{Bashinsky:2003tk},
\begin{equation}
\label{eq:Rnu}
    R_\nu = \dfrac{\rho_\nu^\mathrm{free}}{\rho_\gamma+\rho_\nu}\,.
\end{equation}
Here, $\rho_\gamma$ and $\rho_\nu$ are the total photon and neutrino energy densities, respectively, and $\rho_\nu^\mathrm{free}$ is the total free streaming neutrino density. Clearly, this ratio is maximum in \lcdm{} when all neutrinos are free streaming. Whereas in \three{}, these changes are almost absent as all three flavors are strongly interacting.  The effects are intermediate in \two{} and \one{} where the neutrinos are partially free streaming. 
These will be demonstrated in the results in the next section. Finally, we emphasize that even though we consider only neutrinos as the free-streaming species in this paper, the same physics applies to any radiation-like species.

\section{Results from ACT and Planck + ACT}
\label{sec:results}
We use the CMB data of ACT from four observing seasons to constrain both the flavor universal and non-universal self-interaction scenarios. The flavor universal case has been previously studied in Ref.\,\cite{Kreisch:2022zxp}. However, our analysis of flavor universal self-interaction differs from the previous study in a few key aspects. In majority of the analyses in Ref.\,\cite{Kreisch:2022zxp}, the authors have varied both the effective number of degrees of freedom in neutrinos $\neff$ and neutrinos mass while deriving constraints of flavor universal self-interaction. Ref.\,\cite{Kreisch:2022zxp} did consider a one-parameter $(\lgeff)$ extension which is similar to our case, however, the neutrino mass was set to a non-zero value.
In our baseline models, we have treated the neutrinos to be massless and fixed the $\neff = 3.046$ according to the Big Bang nucleosynthesis calculation assuming the Standard Model. Therefore, our analysis is the minimal extension of the \lcdm{} scenario where we have added only one new parameter which is the neutrino self-interaction strength. We also study cases where we vary both the $\neff$ and included the effects of neutrino mass.
Whereas, for the flavor non-universal case, we derive the first constraints using the ACT data.

\subsection{Datasets \& Methodology}
In our analysis, we have used the ACT DR4 which includes small-scale measurements up to $\ell = 4126$. ACT is a ground-based telescope whose dataset does not extend to the largest observable scale. Therefore, to get the most stringent constraints, we combined the ACT dataset with the Planck measurement following the prescription in Ref\,\cite{ACT:2020gnv}. Below, we detail all the datasets used in the analysis.
\begin{itemize}
    \item ACT: We use the \texttt{actpollite\_dr4} likelihood code from the ACT collaboration which includes the DR4 data release\,\cite{ACT:2020gnv}. The likelihood contains TT measurement $(600 < \ell < 4126 )$ and TE and EE measurements $(350 < \ell < 4126 )$. Because ACT lacks the large-scale data, it cannot constrain the reionization optical depth $\treio$. Hence, we use a Gaussian prior centered at $\treio=0.06\pm0.01$ while analyzing the ACT data alone\,\cite{ACTPol:2016kmo}.\footnote{This value differs slightly from the value used in Ref.\,\cite{ACT:2020gnv,Kreisch:2022zxp}: $\treio=0.065\pm0.015$}
    
    \item Planck: We use the Planck 2018 high-$\ell$ \texttt{plik\_lite}\footnote{Following earlier analyses, the nuisance parameter marginalized `\texttt{lite}' likelihood is used to reduce the dimensionality of the parameter space for MCMC runs using \texttt{Multinest}.} likelihood containing TT measurements $(30 < \ell < 2508 )$ and TE and EE measurements $(30 < \ell < 1996 )$\,\cite{Planck:2019nip}. We also include the low-$\ell$ TT likelihood which includes TT measurements below $\ell < 29$ and the lensing likelihood.
    
    \item Planck + ACT: When combining ACT with Planck data we removed ACT TT data below $\ell < 1800$ following Ref.\,\cite{ACT:2020gnv}. 
\end{itemize}
We used a modified version of \texttt{CLASS}\,\cite{Audren:2012wb,Das:2020xke}, \texttt{MontePython}\,\cite{Brinckmann:2018cvx} and \texttt{Multinest}\,\cite{2014A&A...564A.125B,Feroz:2007kg,Feroz:2008xx,Feroz:2013hea} for the MCMC runs, and \texttt{GetDist}\,\cite{Lewis:2019xzd} to analyze the chains and plot the posterior distributions.

\subsection{Baseline analysis: Massless neutrinos}\label{sec:baseline}
In our baseline analysis, we have considered the neutrinos to be massless and the $\neff$ is set to $3.046$. We analyzed the three scenarios \three, \two, and \one{} where three, two, and one neutrino species are self-interacting, respectively, among the total three species. Similar to the previous analysis we have used the following priors in Table\,\ref{tab:baseline_prior} for baseline analysis.
\renewcommand*{\arraystretch}{1.8}
\begin{table}[t]
    \centering
    \begin{tabular}{|c|>{\centering\arraybackslash}p{3cm}|}
    \hline
        Parameters &  Prior\\
    \hline
        $10^{2}\omega_{b }$ & $[1.00,4.00]$\\
        
        $\omega_{\rm cdm } $ & $[0.08,	0.16]$\\
        
        $100\theta{}_{s } $ & $[1.03	,1.05]$\\
        
        $\ln 10^{10}A_{s }$ & $[2.0	,4.0]$\\
        
        $n_s$ & $[0.9,	1.02]$ \\
        
        $\tau_{\rm reio }$ & $[0.004,	0.25]$ \\
        
        $\log_{10} [{\rm G_{eff}} / {\rm MeV}^{-2}]$ & $[-5.0	,-0.8]$\\
        \hline
        
    \end{tabular}
    \caption{Prior ranges for primary cosmological parameters for the baseline analysis. Although ACT data extends to higher multipole, $\lgeff = -5.0$ still corresponds to the mode for which neutrinos decouple before the horizon entry of the smallest scale probed by ACT (see Fig.2 of Ref.~\cite{Das:2020xke}). For $\log_{10}G_\mathrm{eff}\simeq -5$ neutrinos would decouple at a redshift much higher than $10^5$, which corresponds to multipoles $\ell>10^4$ in the CMB spectrum well beyond the scales probed by ACT. Thus, $\lgeff = -5.0$ safely represents the $\Lambda$CDM limit even for the ACT dataset.}
    \label{tab:baseline_prior}
\end{table}
\begin{figure}[t]
    \centering
    \includegraphics[width=0.49\textwidth]{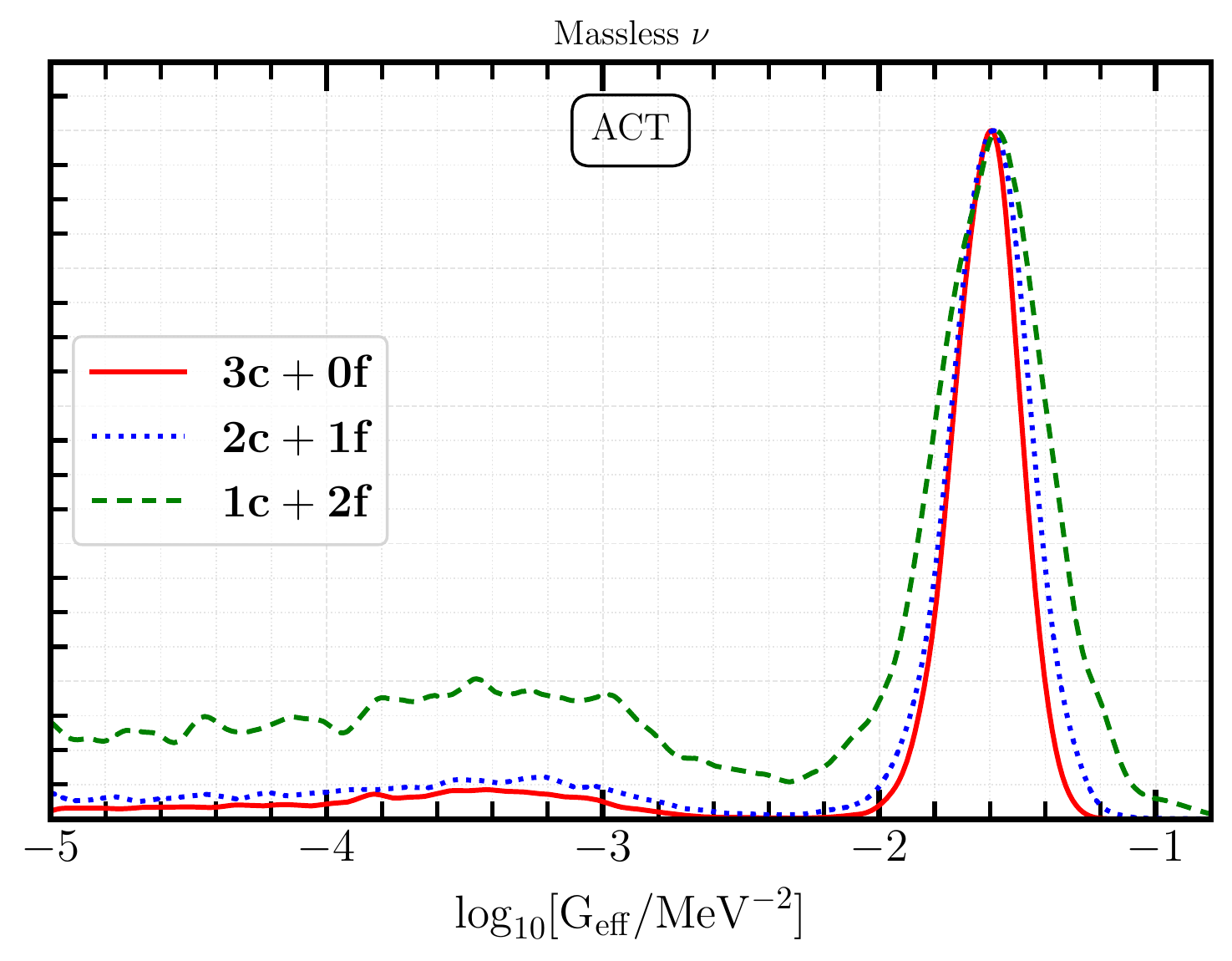}
    \includegraphics[width=0.49\textwidth]{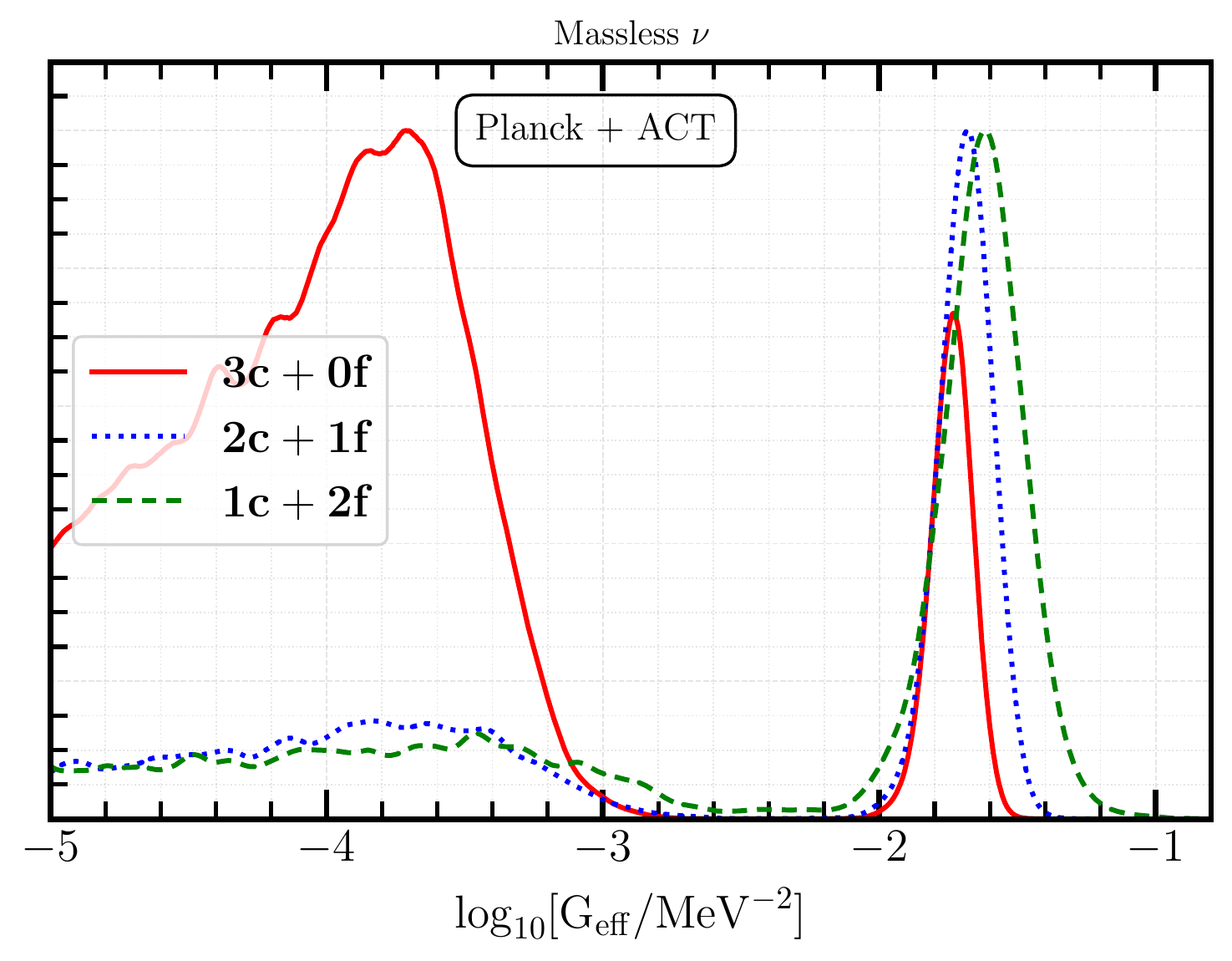}
    \caption{1D posteriors of $\lgeff$ in three coupling scenarios using the  ACT (left) \& ACT+Planck (right) data. The ACT data shows a preference for the SI over MI mode irrespective of the number of coupled neutrino species, which stands in stark contrast with our earlier analysis using Planck 2018 data in Ref.\,\cite{Das:2020xke}. When combined with Planck, the SI mode significance is reduced compared to the MI mode for the flavor-universal case. For flavor-specific cases, SI mode yields a better fit to the data. Additionally, the SI mode in the ACT-only analysis favors a slightly larger value of $\lgeff$.}
    \label{fig:geff-act}
\end{figure}

Fig.\,\ref{fig:geff-act} summarizes the main results of this work for the massless neutrino self-interaction. As in earlier works, we find two separate modes in the $\lgeff$ posterior distribution. The strongly interacting (SI) mode peaks at a higher value of $\lgeff$, and the moderately interesting (MI) mode shows up at a comparatively weaker value closer to the \lcdm{} limit. The ACT data alone prefers the SI mode over the MI mode for all the scenarios. This preference for strong interaction has also been observed in the flavor-universal case (similar to the \three{} scenario) studied in Ref.\,\cite{Kreisch:2022zxp}. This can be attributed to the better fit to the ACT EE data by the SI mode in the $700\lesssim\ell\lesssim1000$ range of multipoles\,\cite{Kreisch:2022zxp}. 

In the flavor non-universal scenarios, namely the \two{} and \one{} cases, the relative significance of the MI mode relative to the SI mode is higher. In the flavor non-universal cases, the changes in the CMB spectrum corresponding to the MI and SI points are smaller compared to the flavor-universal case because a significant portion of the total radiation is simply free-streaming in non-universal case\,\cite{Das:2020xke}. This results in a smaller difference in $\chi^2$ between the best points of the two modes and, thus, results in a smaller peak height difference in the posterior plot corresponding to the two modes.

Moreover, the SI mode peak is slightly shifted to higher $\lgeff$ compared to the Planck-only result (see Fig.\,1 in Ref.\,\cite{Das:2020xke}). Thus ACT data prefer a delayed neutrino decoupling compared to the Planck data. In the right panel of Fig.\,\ref{fig:geff-act} we show the results of the combined analysis with the ACT and Planck data. The addition of ACT data with Planck increases the significance of the SI mode for all cases. Particularly for the flavor non-universal cases, the SI mode provides a better fit to the combined data. Since Planck prefers a smaller value of $\lgeff$ for the SI mode, in the combined analyses, the SI mode peak positions are stretched toward a slightly smaller value compared to the ACT result.

\begin{figure}[t]
    \centering
    \includegraphics[width=0.7\textwidth]{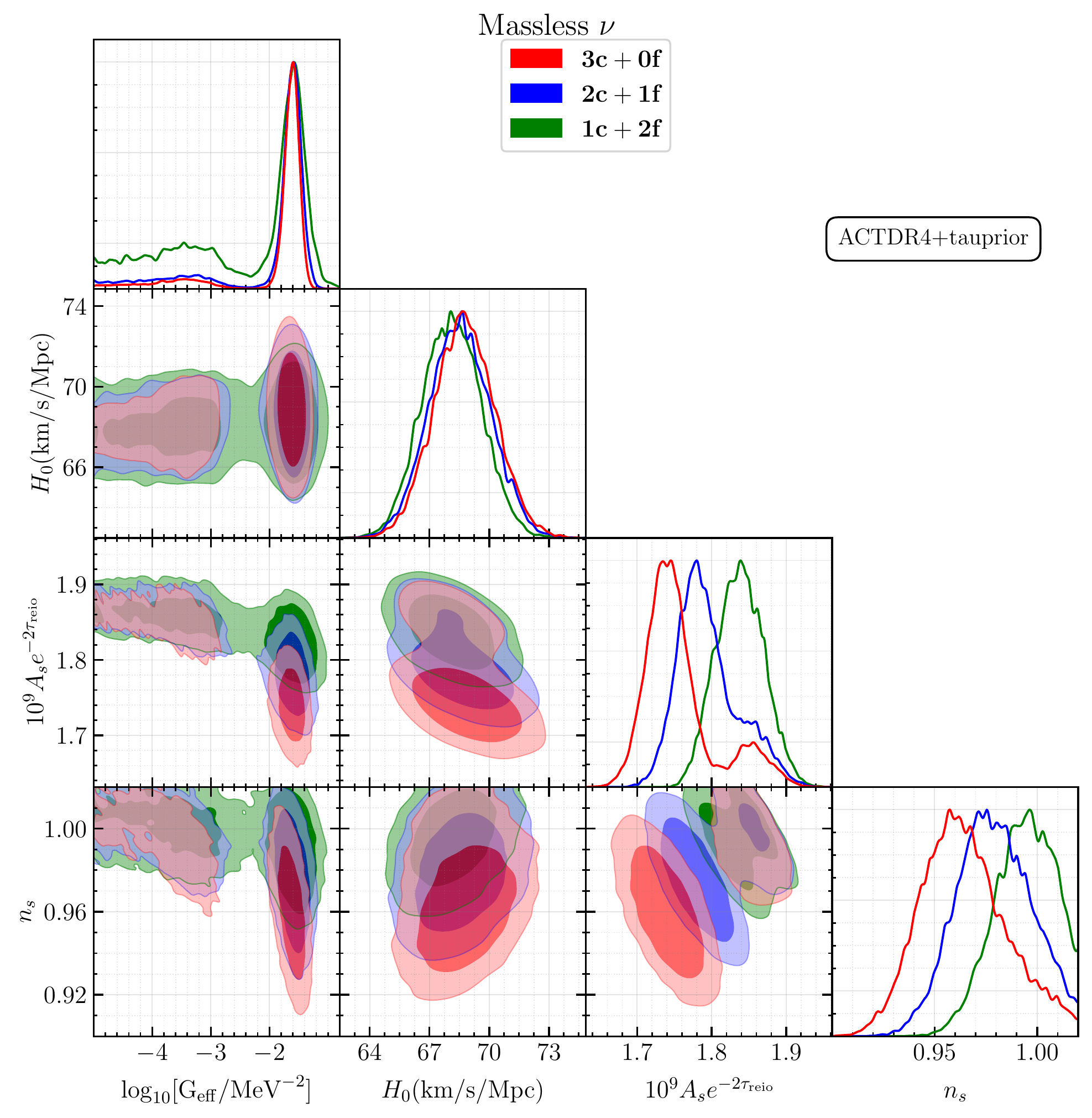}
    \caption{The posterior distributions of $\lgeff, H_0, A_se^{-2\treio}$, and $n_s$ using the ACT data alongwith the $\tau$-prior. Note the large boost in the significance of the SI mode in the $\lgeff$ posterior. See Fig.\,\ref{fig:triall_baseline_ACT} in Appendix.\,\ref{app:triplots} for an extended parameter plot.}
    \label{fig:tri-act}
\end{figure}

\begin{figure}[t]
    \centering
    \includegraphics[width=0.7\textwidth]{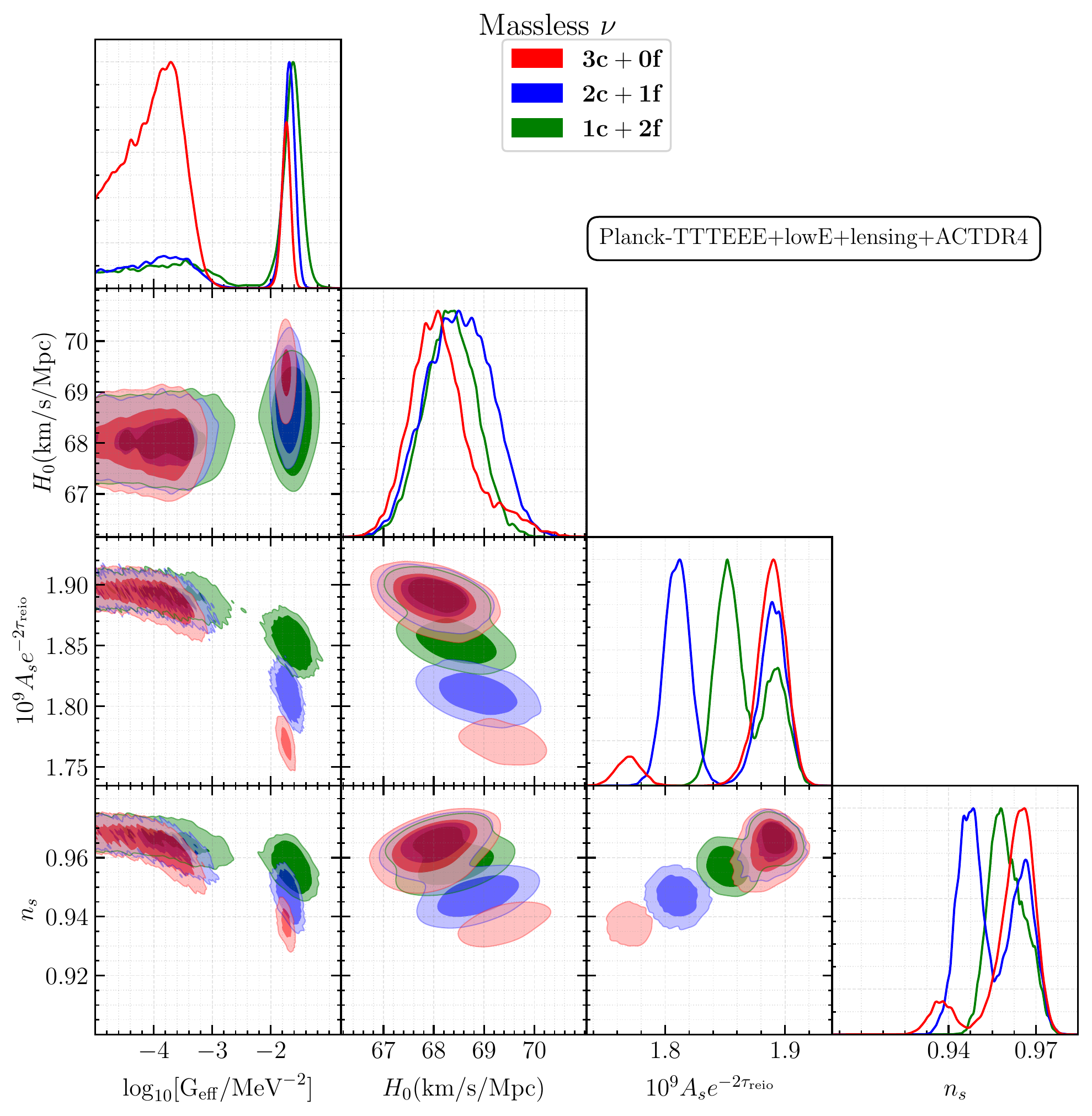}
    \caption{The posterior distributions of $\lgeff, H_0, A_se^{-2\treio}$, and $n_s$ using the combined ACT and Planck data. See Fig.\,\ref{fig:triall_baseline} in Appendix.~\ref{app:triplots} for an extended parameter plot.}
    \label{fig:tri-act-planck}
\end{figure}

In Fig.\,\ref{fig:tri-act} and \ref{fig:tri-act-planck}, we show the triangle plots for some relevant cosmological parameters using ACT and ACT+Planck data, respectively. For both datasets, there is an expected shift of the parameters $A_se^{-2\tau_\mathrm{reio}}$ and $n_s$ toward larger values for the SI mode when the number of interacting neutrino species is increased, which can be clearly seen from the 2D plots. As explained in Ref.\,\cite{Das:2020xke}, this is because the scale-dependent power enhancement from large $\geff$ depends on the total number of interacting neutrinos. This shift is apparent from Fig.\,\ref{fig:tri-act} as for the ACT analysis the posterior is entirely dominated by the SI mode. However, in the Planck + ACT analysis in Fig.~\ref{fig:tri-act-planck}, the MI mode becomes dominant in \three{}. This leads to the bigger peak to the right in the $A_se^{-2\tau_\mathrm{reio}}$ and $n_s$ posteriors which should not be confused with the dominant SI mode peaks for \two{} and \one{}. In fact, it is easier to identify the individual modes (SI/MI) for any parameter by comparing with the 2D posterior with $\lgeff$.


To study the individual modes in detail, we performed a separate analysis where we fixed the prior range on $\lgeff$ to separate out the MI and SI mode according to Table\,\ref{tab:prior_misi}\,\cite{Das:2020xke}.
\begin{table}[t]
    \centering
    \begin{tabular}{|c|c|}
    \hline
       Mode  &  Prior range on $\lgeff$\\
    \hline
       MI  & $[-5.0, -2.7]$ \\
       SI & $[-2.7, -0.8]$\\
       \hline
    \end{tabular}
    \caption{Prior ranges for $\lgeff$ for MI and SI mode separation. The ranges for other cosmological parameters remain the same as in Table\,\ref{tab:baseline_prior}.}
    \label{tab:prior_misi}
\end{table}
The parameter values and limits of the cosmological parameters for the ACT and Planck + ACT datasets are quoted in Table\,\ref{tab:act_baseline} and \ref{tab:act-planck_baseline}, respectively. In the tables, we also quote the $\chi^2$ difference for each of the models from the corresponding 6-parameter $\Lambda$CDM model. We also quote the Akaike Information Criterium (AIC) criterion defined as\,\cite{1100705},
\begin{equation}
    \Delta{\rm AIC} = \chi^2_{{\rm min},\mathcal{M}} - \chi^2_{{\rm min},\Lambda{\rm CDM}} + 2(N_\mathcal{M} - N_{\Lambda{\rm CDM}}).
\end{equation}
Here $\chi^2_{{\rm min},i}$ and $N_i$ are the minimum $\chi^2$ and number of parameters for $i$ model, respectively. Negative values $\Delta{\rm AIC}$ signify the preference of the models over $\Lambda$CDM. Additionally, for the baseline model, we also compute the Bayes factor $(\mathcal{B}_{\rm SI})$ which is the ratio of Bayesian evidence of the SI over the MI mode.
\begin{equation}
    \mathcal{B}_{\rm SI} = {\mathcal{Z}_{\rm SI} \over \mathcal{Z}_{\rm MI}}
\end{equation}
The Bayesian evidence which is defined below is computed using \texttt{MultiNest} algorithm.
    \begin{equation}
    \mathcal{Z}_M \equiv \mathrm{Pr}(\mathrm{d}|M) = \int \mathrm{Pr}(\mathrm{d}|\theta,M) \mathrm{Pr}(\theta|M) \mathrm{d}\theta\,,
    \end{equation}
    where $\mathrm{d}$ is the data and $\theta$ are the cosmological parameters for model $M$. $\mathcal{B}_\mathrm{SI} > 1$ signifies that the SI mode is more significant than the MI mode. In Table\,\ref{tab:bayes}, we show the Bayes for all the self-interaction scenarios for the baseline model.

For the ACT dataset, both $\Delta$AIC and $\mathcal{B}_{\rm SI}$ show that more interacting flavors provide a better fit to the data compared to $\Lambda$CDM. Note that, the \three{} model has the most negative $\Delta$AIC, however, \two{} has the largest Bayes factor\footnote{This is not surprising because these two methods measure different quantities for model comparison. The Bayes factor is related to the ratio of areas under the SI and MI modes in the 1D posterior, respectively. We can see from Fig.\,\ref{fig:geff-act} that visibly the area under the SI mode of \two{} is greater than \three{} which can offset the slight increase in area under the MI mode.}. Since ACT data lacks large-scale information, the changes induced by the strong neutrino self-interaction can be relatively easily compensated by modifying other parameters, most notably, $A_se^{-2\tau_\mathrm{reio}}$ and $n_s$ as can be seen from the 2D panels in Fig.\,\ref{fig:tri-act}. While analyzing in combination with the Planck data, this freedom is taken away and it has been established by previous studies that \three{} scenario does not provide a good fit to the Planck data\,\cite{Das:2020xke}. Therefore, for Planck + ACT results, a lesser amount of self-interacting neutrino is favored. This shows that Planck has more statistical power (more precision) than the ACT data as the combined analysis prefers the MI mode in the \three{} case. While for \two{} and \one{} both Planck and ACT have a stronger preference for the SI mode compared to the flavor universal case which results in a greater significance of the SI mode over the MI one compared to the \three{} scenario.

\begin{table}[t]
    \resizebox{\textwidth}{!}{
    \begin{tabular}{|c|P{2.68cm}P{2.68cm}|P{2.68cm}P{2.68cm}|P{2.68cm}P{2.68cm}|}
\hline
Massless $\nu$ &\multicolumn{2}{c}{$\mathbf{3c+0f}$}\vline &\multicolumn{2}{c}{$\mathbf{2c+1f}$}\vline &\multicolumn{2}{c}{$\mathbf{1c+2f}$}\vline\\
\hline
Parameters  & MI & SI & MI & SI & MI & SI\\
\hline
$10^2 \omega_{\rm b}$ & $ 2.151\pm 0.030$ & $ 2.153\pm 0.031$ & $ 2.154\pm 0.030$ & $ 2.153\pm 0.030$ & $ 2.155\pm 0.029$ & $ 2.155\pm 0.030$\\
$\omega_{\rm cdm}$ & $ 0.1190\pm 0.0037$ & $ 0.1215\pm 0.0039$ & $ 0.1192\pm 0.0037$ & $ 0.1205\pm 0.0039$ & $ 0.1194\pm 0.0037$ & $ 0.1199\pm 0.0038$\\
$100\theta{}_{s }$ & $ 1.04358\pm 0.00076$ & $ 1.04872^{+0.00084}_{-0.00058}$ & $ 1.04345\pm 0.00073$ & $ 1.04684\pm 0.00079$ & $ 1.04335\pm 0.00070$ & $ 1.04495\pm 0.00076$\\
$\ln10^{10}A_s$ & $ 3.038\pm 0.024$ & $ 2.974\pm 0.024$ & $ 3.042\pm 0.023$ & $ 2.998\pm 0.024$ & $ 3.045\pm 0.023$ & $ 3.022\pm 0.023$\\
$n_{s }$ & $ 0.991^{+0.019}_{-0.013}$ & $ 0.960^{+0.016}_{-0.018}$ & $ 0.996^{+0.017}_{-0.011}$ & $ 0.975\pm 0.016$ & $ 0.999^{+0.016}_{-0.0094}$ & $ 0.990\pm 0.015$\\
$\tau_{\rm reio}$ & $ 0.0596\pm 0.0097$ & $ 0.0588\pm 0.0098$ & $ 0.0600\pm 0.0098$ & $ 0.0594\pm 0.0097$ & $ 0.0600\pm 0.0099$ & $ 0.0598\pm 0.0097$\\
$\log_{10} [{\rm G_{eff}} / {\rm MeV}^{-2}]$ & $ -3.72^{+0.79}_{-0.41}$ & $ -1.62^{+0.14}_{-0.10}$ & $ -3.77\pm 0.60$ & $ -1.62^{+0.17}_{-0.13}$ & $ -3.79^{+0.87}_{-0.75}$ & $ -1.65^{+0.29}_{-0.18}$\\
\hline
$H_0 ({\rm km/s/Mpc})$ & $ 68.1\pm 1.5$ & $ 68.9\pm 1.5$ & $ 68.0\pm 1.5$ & $ 68.6^{+1.4}_{-1.7}$ & $ 67.9\pm 1.5$ & $ 68.3\pm 1.5$\\
$\sigma_8$ & $ 0.840\pm 0.015$ & $ 0.840\pm 0.016$ & $ 0.840^{+0.015}_{-0.014}$ & $ 0.837\pm 0.016$ & $ 0.840\pm 0.015$ & $ 0.838\pm 0.016$\\
\hline
$\chi^2 - \chi^2_{\Lambda{\rm CDM}}$ & $-2.08$ & $-8.26$ & $-1.34$ & $-6.9$ & $-0.69$ & $-4.07$\\
\hline
$\Delta{\rm AIC}$ & $-0.08$ & $-6.26$ & $0.66$ & $-4.9$ & $1.31$ & $-2.07$\\
\hline
\end{tabular}
    }
    \caption{Mean values and 68\% confidence limits for ACT dataset for the baseline massless $\nu$ scenario. We also show the $\chi^2$ difference from the bestfit 6-parameter $\Lambda$CDM model and the corresponding $\Delta$AIC. Two important derived parameters $H_0$ and $\sigma_8$ are also shown. In all tables, the varied and derived parameters are separated by a horizontal line unless mentioned otherwise.}\label{tab:act_baseline}
\end{table}

Another interesting feature of the analysis with the ACT data is the values of $H_0$ for the various scenarios. Historically, flavor-universal neutrino self-interaction was proposed as a solution to the Hubble tension, which is the discrepancy between the CMB and local measurement of $H_0$ using the distance ladder, currently standing at $\sim\,4.4\sigma$\,\cite{Riess:2021jrx}. However, with the release of Planck 2018 polarization data, the overall Planck data predominantly favors the MI mode (which is the $\Lambda$CDM limit) over the SI mode\,\cite{RoyChoudhury:2020dmd,Das:2020xke}. In the flavor non-universal case, although the significance of the SI mode peak increases dramatically, the corresponding $H_0$ value does not increase appreciably to resolve the tension\,\cite{Das:2020xke}. The changes in the CMB spectra compared to $\Lambda$CDM are milder in flavor-specific neutrino self-interaction compared to the flavor universal case as only a smaller fraction of the total radiation is self-interacting. Therefore, the shifts in other cosmological parameters to compensate for these changes (to arrive at a good fit to the data) are also smaller in flavor-specific cases. This is the reason for a smaller upward shift of the $H_0$ for flavor non-universal cases. This is a robust prediction that also holds true when the effective number of neutrinos and neutrino mass are varied.

\begin{figure}
    \centering 
    \includegraphics[width=0.49\textwidth]{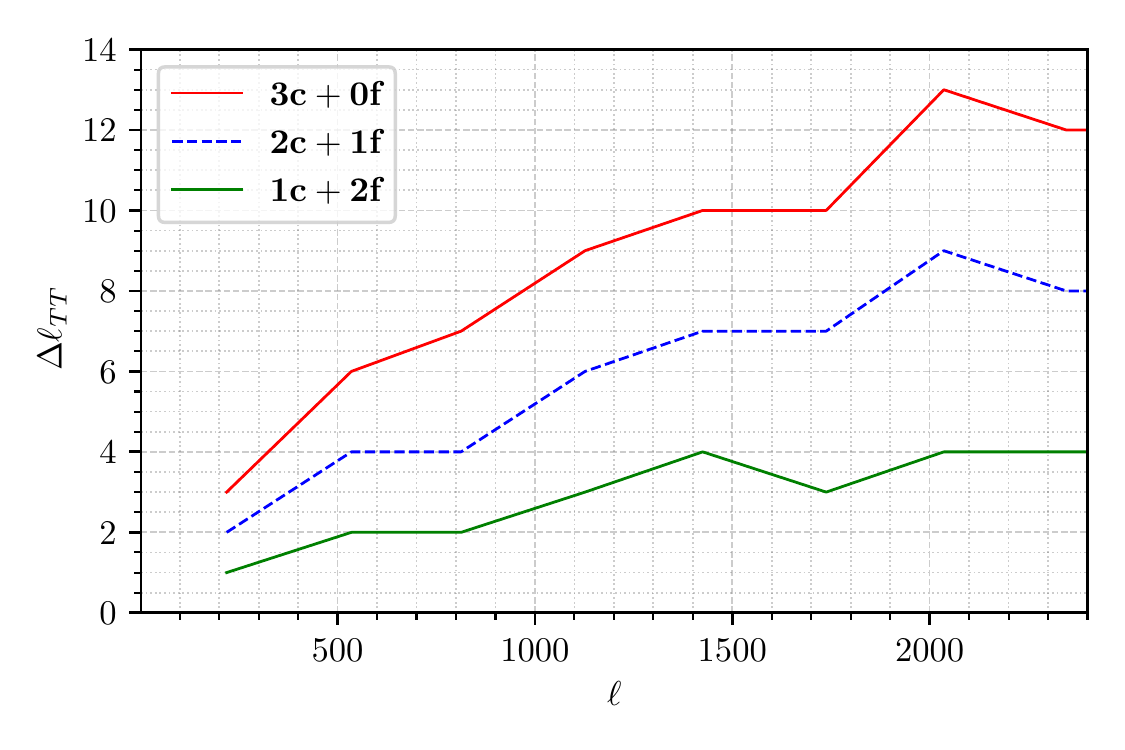}
    \includegraphics[width=0.49\textwidth]{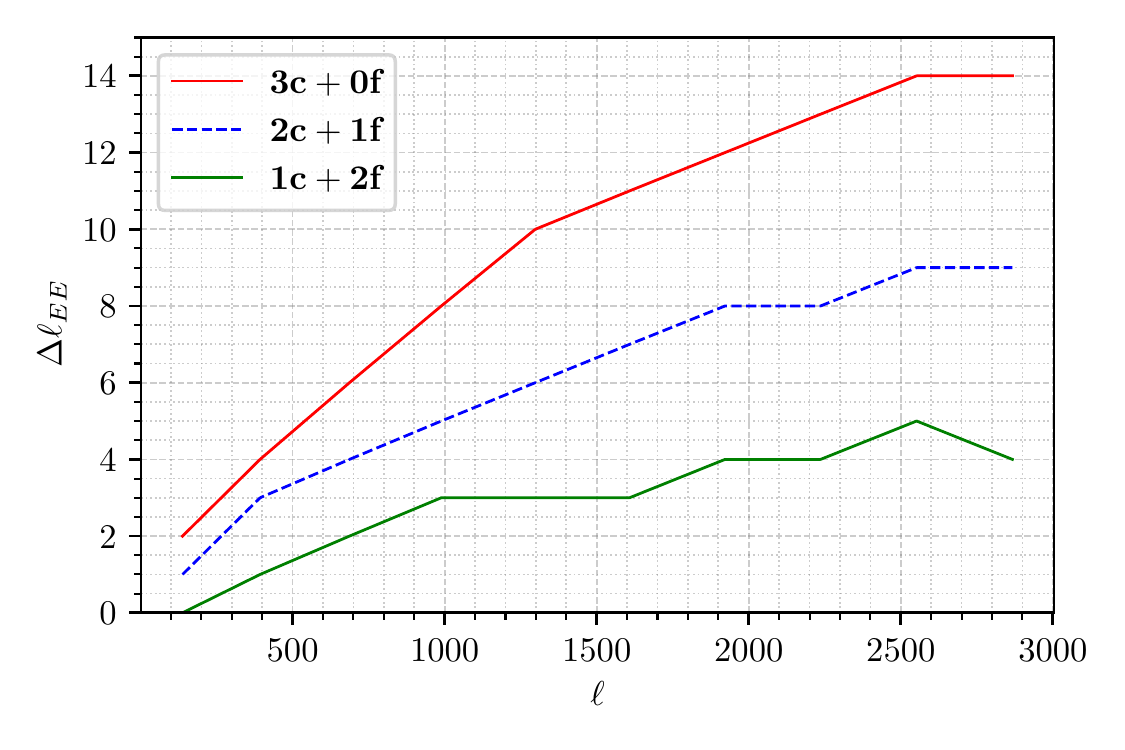}
    \caption{Shift in the CMB acoustic peak positions compared to $\Lambda$CDM for the TT (Left) and EE (Right) spectra for $\lgeff=-1.0 $. The origin of the peak shift is the change of the neutrino-induced phase-shift in the CMB acoustic peaks due to neutrino-self interaction~\cite{Bashinsky:2003tk,Ghosh:2019tab,Baumann:2015rya}. The phase shift, hence the peak shift, is proportional to the amount of coupled neutrino flavor. The phase shift is the primary driver for the increase in $H_0$ in the neutrino-self interaction scenarios, specifically when $N_{\rm eff}$ is kept fixed.}
    \label{fig:peak_shift}
\end{figure}

The increase in $H_0$ in our baseline scenario results from the modification of the acoustic phase shift.
Free streaming neutrinos induce a scale-dependent phase shift on photon-baryon acoustic oscillation which moves the CMB peaks toward smaller multipole. Due to the self-interaction, neutrinos stop free-streaming and the acoustic phase shift is modified. The shift in the CMB peak position due to the reduction of the phase shift can be compensated by the increase of the $H_0$\,\cite{Ghosh:2019tab}. In Fig.~\ref{fig:peak_shift}, we show the shifts of the CMB peaks for the TT and EE spectra for flavor dependent neutrino self-interaction.
This is the primary mechanism for the increase of $H_0$ when $N_{\rm eff}$ is not allowed to vary. The amount of neutrino-induced phase follows the scaling relation:
\begin{equation}
    \phi_\nu \propto   R_{\nu} = R_{\nu}^\text{\lcdm{}} \times \begin{cases}
  0,~~\,\qquad {\rm for~}\text{\three}\\[1.5ex] {\Large \sfrac{1}{3}},\qquad {\rm for~}\text{\two} \\[1.5ex] {\Large\sfrac{2}{3}},\qquad {\rm for~}\text{\one}
  \end{cases}  
\end{equation}
Here $R_{\nu}$ is the free-streaming neutrino energy fraction defined in Eq.(\ref{eq:Rnu}) and $\phi_\nu$ is the acoustic phase induced by neutrinos. The proportionality of the phase shift with the number of interacting neutrino flavors can be clearly seen in Fig.\,\ref{fig:peak_shift}. A tale-tell signature of this mechanism is a shift in the parameter $\theta_s$ which is the angular size of the sound horizon at recombination\,\cite{Ghosh:2019tab,Das:2020xke}:
\begin{equation}
    \label{eq:thetas_change}
    \frac{\Delta \theta_s}{\theta_s} \propto \Delta \phi\;,
\end{equation}
where $\Delta\phi$ is the change of the neutrino-induced phase compared to $\Lambda$CDM\footnote{Since the MI mode is the \lcdm{} limit for neutrino self interaction, $\Delta \theta_s = \theta_s^{\rm (SI)} - \theta_s^{(\Lambda\rm{CDM} )} \approx \theta_s^{\rm (SI)} - \theta_s^{(\rm MI)}$ .}.
Thus the following scaling relation is a robust prediction for the flavor-specific scenario,
\begin{equation}
\label{eq:thetas_scaling}
    \Delta \theta_s|_\text{\one} \approx \frac{1}{2}\Delta \theta_s|_\text{\two} \approx \frac{1}{3}\Delta \theta_s|_\text{\three}.
\end{equation}

One can check the validity of Eq.(\ref{eq:thetas_scaling}) from Table\,\ref{tab:act_baseline} and \ref{tab:act-planck_baseline}\footnote{While comparing with numbers from the tables $\Delta \theta_s$ is the difference between the MI and SI mode \emph{mean} values.}. Later in the paper, we will present results for varying $\neff$ of both coupled and free-streaming nature. The above equations also hold true in those cases by the appropriate modification of $R_{\nu}$.

However, a sizable positive change in $\Delta \theta_s$ does not always translate to a proportionately higher value of the $H_0$. The exact relation between $\theta_s$ and $H_0$ involves other cosmological parameters such as baryon and dark matter energy densities. As the ACT data lack large-scale information, these energy densities are less constrained, resulting in a slightly weaker increase in $H_0$ in the ACT case. As the modification of $\theta_s$ is the primary driver for $H_0$ increase for our baseline model, a similar scaling in terms of $H_0$ also holds true,
\begin{equation}
\label{eq:H0_scaling}
    \Delta H_0|_\text{\one} \approx \frac{1}{2}\Delta H_0|_\text{\two} \approx \frac{1}{3}\Delta H_0|_\text{\three}.
\end{equation}






\begin{table}[t]
    \resizebox{\textwidth}{!}{
    \begin{tabular}{|c|P{2.68cm}P{2.68cm}|P{2.68cm}P{2.68cm}|P{2.68cm}P{2.68cm}|}
\hline
Massless $\nu$ &\multicolumn{2}{c}{$\mathbf{3c+0f}$}\vline &\multicolumn{2}{c}{$\mathbf{2c+1f}$}\vline &\multicolumn{2}{c}{$\mathbf{1c+2f}$}\vline\\
\hline
Parameters  & MI & SI & MI & SI & MI & SI\\
\hline
$10^2 \omega_{\rm b}$ & $ 2.232\pm 0.013$ & $ 2.245^{+0.012}_{-0.014}$ & $ 2.233\pm 0.012$ & $ 2.239\pm 0.013$ & $ 2.235\pm 0.012$ & $ 2.236\pm 0.013$\\
$\omega_{\rm cdm}$ & $ 0.1197\pm 0.0011$ & $ 0.1213\pm 0.0011$ & $ 0.1197\pm 0.0011$ & $ 0.1208\pm 0.0011$ & $ 0.1197\pm 0.0011$ & $ 0.1203\pm 0.0011$\\
$100\theta{}_{s }$ & $ 1.04210\pm 0.00029$ & $ 1.04738^{+0.00044}_{-0.00038}$ & $ 1.04211\pm 0.00029$ & $ 1.04560^{+0.00040}_{-0.00034}$ & $ 1.04209^{+0.00026}_{-0.00029}$ & $ 1.04382^{+0.00038}_{-0.00031}$\\
$\ln10^{10}A_s$ & $ 3.048\pm 0.012$ & $ 2.981\pm 0.012$ & $ 3.049\pm 0.012$ & $ 3.004\pm 0.012$ & $ 3.051\pm 0.012$ & $ 3.029\pm 0.012$\\
$n_{s }$ & $ 0.9639^{+0.0061}_{-0.0044}$ & $ 0.9373\pm 0.0042$ & $ 0.9646^{+0.0057}_{-0.0041}$ & $ 0.9469\pm 0.0043$ & $ 0.9661\pm 0.0042$ & $ 0.9569\pm 0.0041$\\
$\tau_{\rm reio}$ & $ 0.0549\pm 0.0057$ & $ 0.0539\pm 0.0058$ & $ 0.0551\pm 0.0059$ & $ 0.0542\pm 0.0059$ & $ 0.0554^{+0.0052}_{-0.0061}$ & $ 0.0548\pm 0.0058$\\
$\log_{10} [{\rm G_{eff}} / {\rm MeV}^{-2}]$ & $ -4.05^{+0.62}_{-0.40}$ & $ -1.737^{+0.086}_{-0.064}$ & $ -3.94\pm 0.53$ & $ -1.69^{+0.11}_{-0.090}$ & $ -3.87^{+0.83}_{-0.69}$ & $ -1.65^{+0.18}_{-0.12}$\\
\hline
$H_0 ({\rm km/s/Mpc})$ & $ 68.01\pm 0.49$ & $ 69.38\pm 0.49$ & $ 68.04\pm 0.48$ & $ 68.87\pm 0.49$ & $ 68.05\pm 0.48$ & $ 68.44\pm 0.48$\\
$\sigma_8$ & $ 0.8291\pm 0.0054$ & $ 0.8335\pm 0.0062$ & $ 0.8289\pm 0.0053$ & $ 0.8293\pm 0.0062$ & $ 0.8286\pm 0.0053$ & $ 0.8270\pm 0.0060$\\
\hline
$\chi^2 - \chi^2_{\Lambda{\rm CDM}}$ & $-1.89$ & $-1.21$ & $-1.41$ & $-5.25$ & $-0.79$ & $-5.17$\\
\hline
$\Delta{\rm AIC}$ & $0.11$ & $0.79$ & $0.59$ & $-3.25$ & $1.21$ & $-3.17$\\
\hline
\end{tabular}
    }
    \caption{Mean values and 68\% confidence limits for the Planck + ACT dataset. We also show the $\chi^2$ difference from the bestfit 6-parameter $\Lambda$CDM model and the corresponding $\Delta$AIC. }\label{tab:act-planck_baseline}
\end{table}

\begin{table}[b]
    \centering
    \resizebox{0.85\textwidth}{!}{
    \begin{tabular}{|l|P{2.68cm}P{2.68cm}P{2.68cm}|}
\hline
Massless $\nu$ & $\mathbf{3c+0f}$ & $\mathbf{2c+1f}$ & $\mathbf{1c+2f}$\\
\hline
ACTDR4+tauprior & $6.576$ & $6.706$ & $2.519$\\
\hline
Planck-TTTEEE+lowE+lensing & $0.024$ & $0.179$ & $0.529$\\
\hline
Planck-TTTEEE+lowE+lensing+ACTDR4 & $0.161$ & $1.291$ & $2.227$\\
\hline
\end{tabular}}
    \caption{Bayes factors for SINU for all datasets.}
    \label{tab:bayes}
\end{table}

\begin{figure}[t]
    \centering 
    \includegraphics[width=0.49\textwidth]{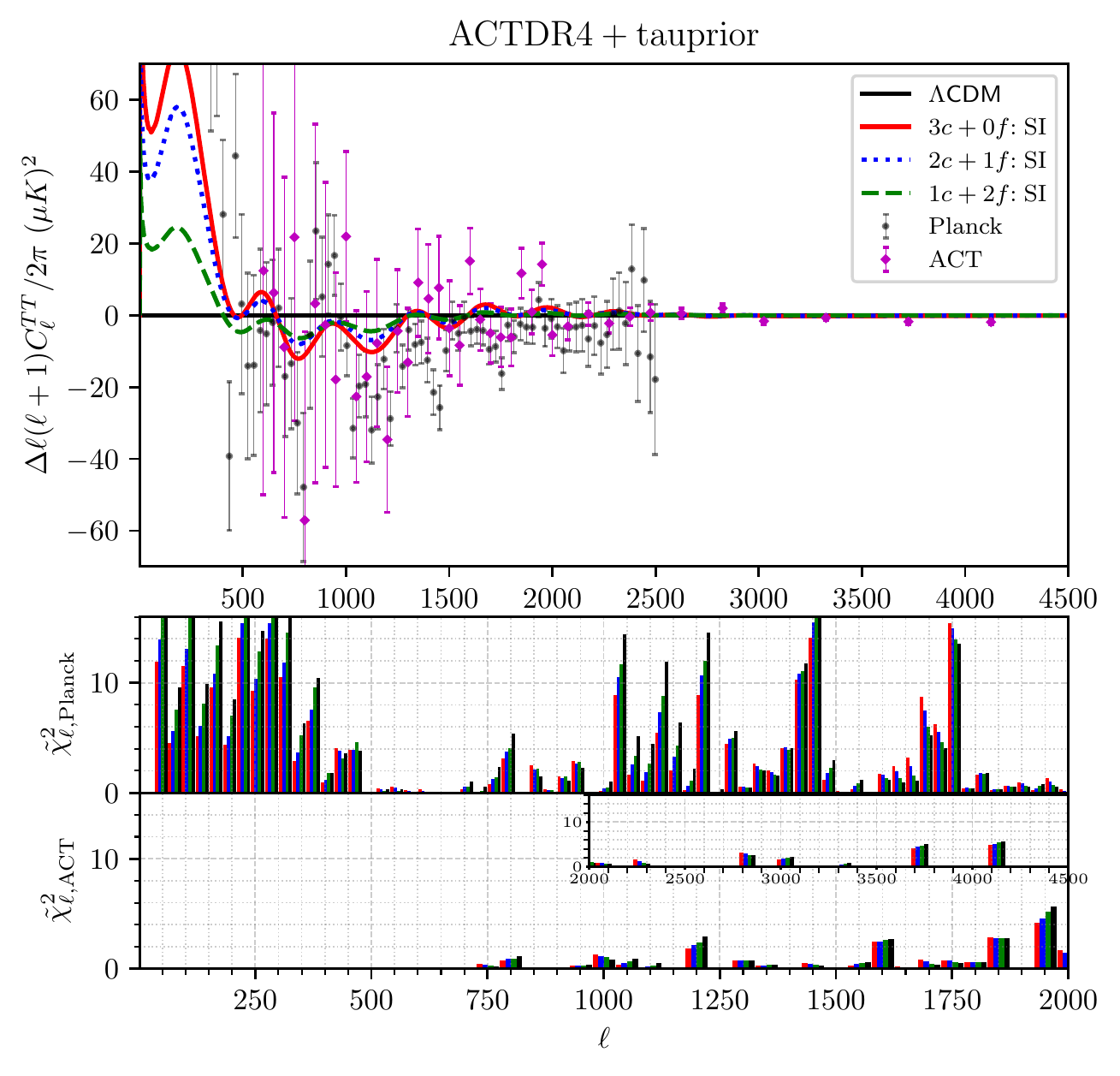}
    \includegraphics[width=0.49\textwidth]{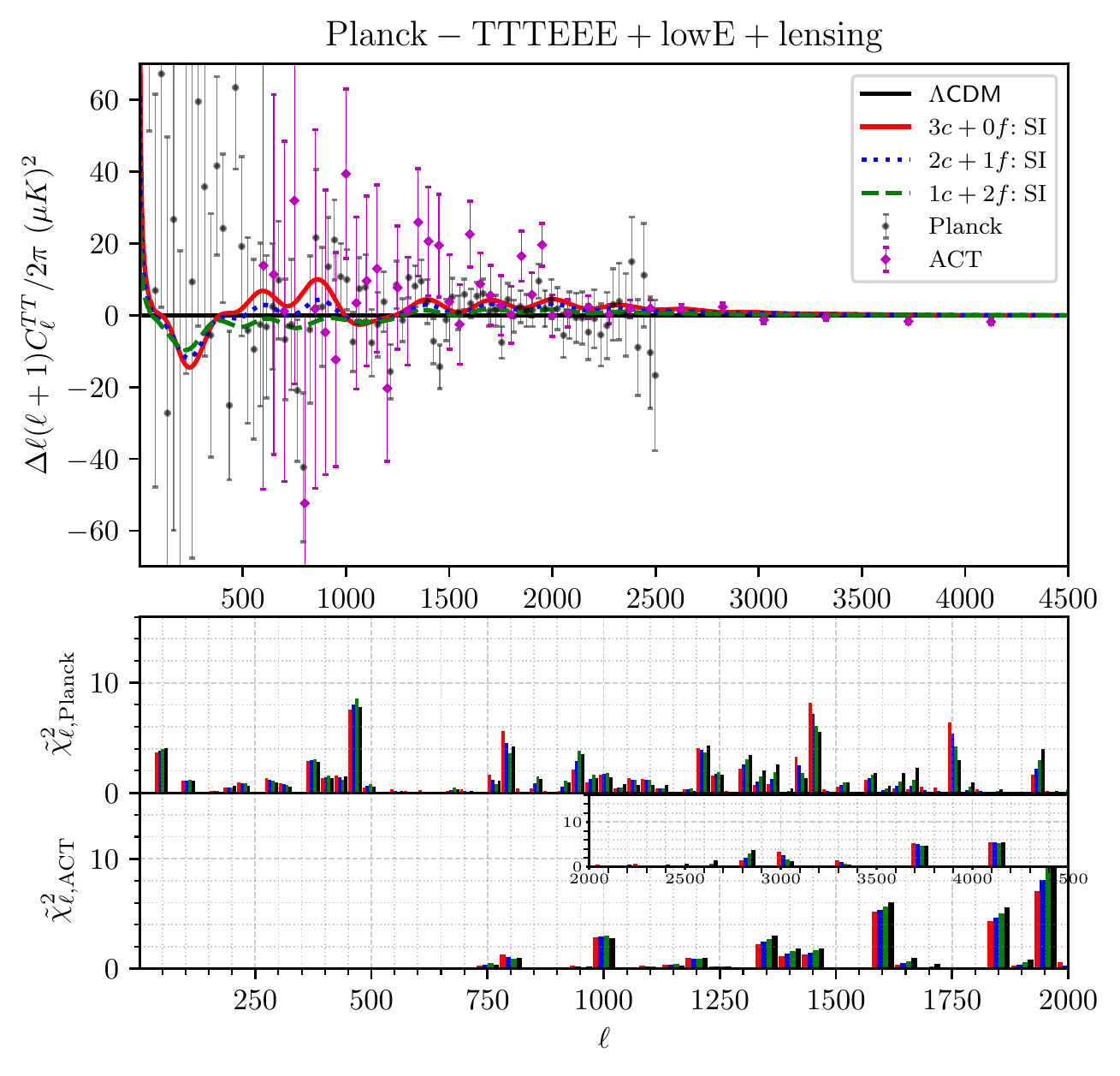}
    \caption{(Top) TT residuals of the bestfit models relative to the corresponding \lcdm{} analyses for the ACT (left) and Planck (right) datasets. (Bottom) The multipole distribution of the $\chi^2_\ell$ for the two bestfit models.}
    \label{fig:residuals_tt}
\end{figure}
\begin{figure}[t]
    \centering 
    \includegraphics[width=0.49\textwidth]{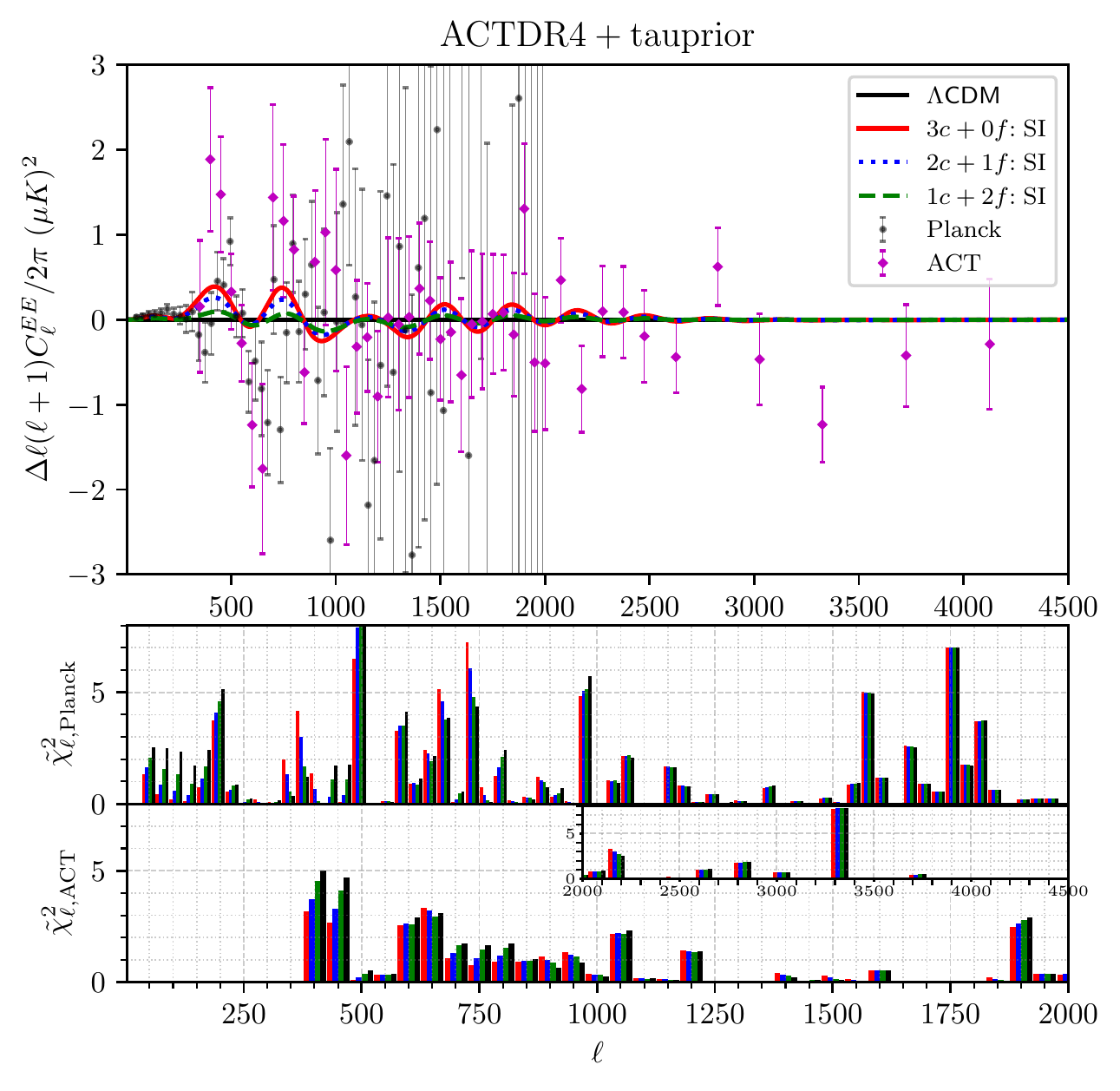}
    \includegraphics[width=0.49\textwidth]{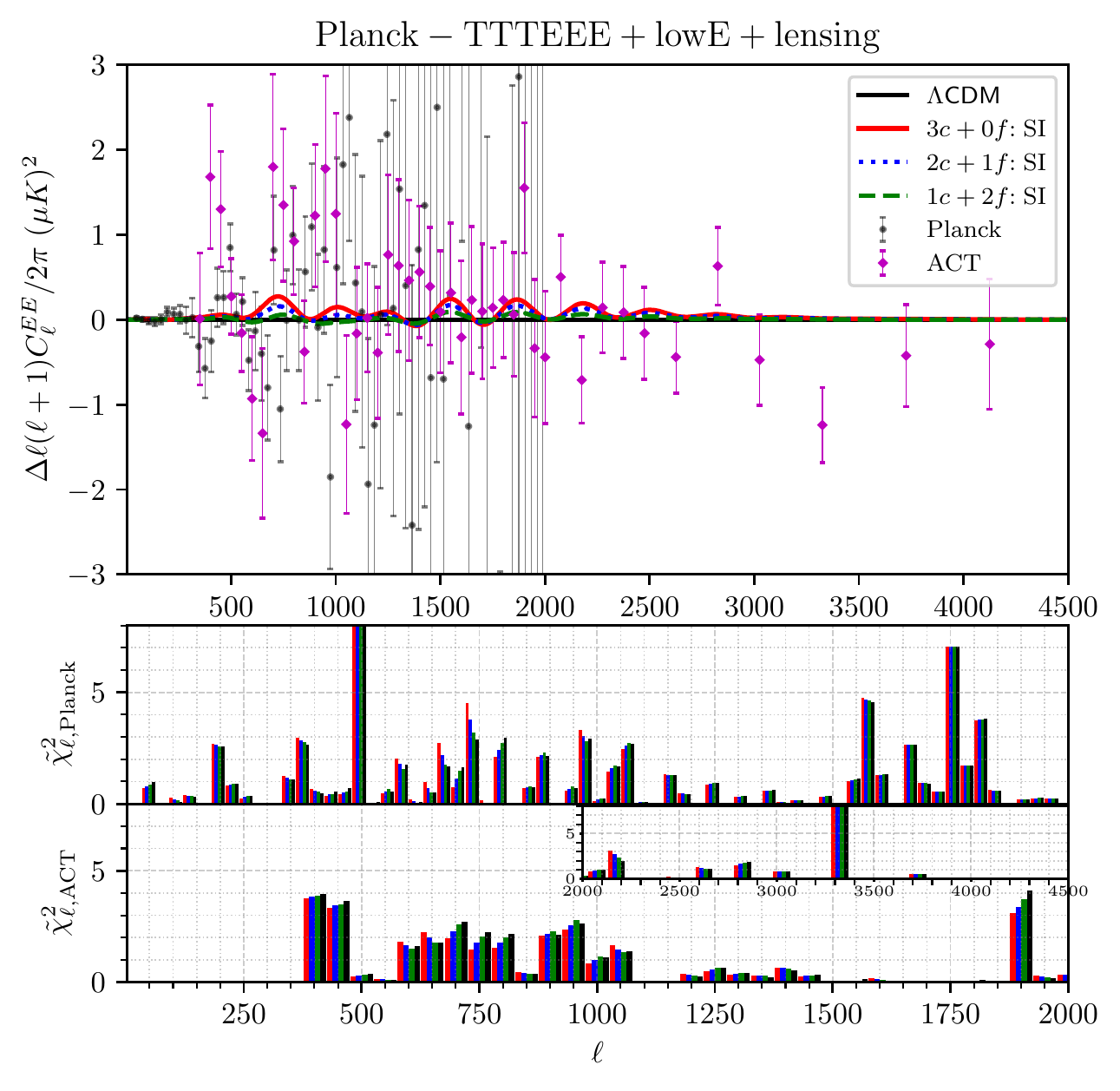}
    \caption{Same as Fig.\,\ref{fig:residuals_tt} for EE spectrum.}
    \label{fig:residuals_ee}
\end{figure}

In Fig.\,\ref{fig:residuals_tt} and \ref{fig:residuals_ee} we show the residuals of the bestfit SINU models w.r.t. the corresponding \lcdm{} models for both ACT (left) and Planck (right) for the TT and EE modes, respectively. To analyze the consistency between the two datasets, we plot the residual with respect to both these datasets in each plot. For better visualization of the residuals, we plot the effective $\chi^2$ for each dataset bin ignoring the correlation between adjacent bins,
\begin{equation}
    \label{eq:eff_chisq}
    \tilde{\chi}^2_{\ell,{\rm Data}} = \frac{(C_{\ell,{\rm Data}} - C_{\ell,{\rm bf}})^2}{(\Delta C_{\ell,{\rm Data}})^2} \qquad\text{for}~\ell \in \ell_{\rm Data(binned)}.
\end{equation}
We use the binning provided in the official data release by the respective collaborations.

For the ACT TT residual plot (Fig.\,\ref{fig:residuals_tt}, left) we see that SINU provides a better fit to the data than the \lcdm{}. This can be verified from the plot of $\tilde{\chi}^2_{\ell,{\rm ACT}}$ where the \three{} overall gives the least $\tilde{\chi}^2$ across all the bins. However, we can see from the $\tilde{\chi}^2_{\ell,{\rm Planck}}$ panel that Planck data highly disfavor this fit. This discrepancy is particularly very large for $\ell \lesssim 500$ since ACT does not have any data in that multipole range. For the Planck TT residual plot (Fig.\,\ref{fig:residuals_tt}, right) we see that \one{} provides an overall better fit to the among SINU scenarios and is very close to the $\Lambda$CDM fit. This can be clearly seen from the $\tilde{\chi}^2_{\ell,{\rm Planck}}$ panel. Interestingly, from both the $\tilde{\chi}^2_{\ell,{\rm ACT}}$ panels in the figure, we see that in general SINU scenarios give smaller $\tilde{\chi}^2$ than $\Lambda$CDM
for ACT. The overall increase in $\tilde{\chi}^2$ is visibly quite smaller going from ACT bestfit to Planck bestfit for $\tilde{\chi}^2_{\ell,{\rm ACT}}$. Thus ACT TT data prefers 
SINU
over $\Lambda$CDM for the bestfit with ACT runs and also for the bestfit with Planck runs.  Thus for Planck+ACT results, the significance of the SI mode increases owing to the addition of ACT data. 
In the EE residual plots (Fig.\,\ref{fig:residuals_tt}), we see the overall same features as in the TT residual plots. 

We can also comment on the compatibility between ACT and Planck data from the same residual plots. TT residual in Fig.\,\ref{fig:residuals_tt} shows that compared to ACT bestfit (left panel) $\tilde{\chi}^2_{\ell,{\rm ACT}}$ increases substantially in the multipole range $1600<\ell<2000$ for the Planck bestfit (right panel). Whereas, for EE residuals (Fig.\,\ref{fig:residuals_tt}) $\tilde{\chi}^2_{\ell,{\rm ACT}}$ increases significantly in the range $600<\ell<1000$. Therefore, TT and EE data from ACT and Planck have a mild disagreement in those corresponding multipole ranges mentioned above. This perhaps points to some unknown systematics in the intermediate-scale data. Future data from ACT, SPT, and other experiments will be crucial in confirming this.

Note that for SI interaction strength neutrinos decouple around redshift $z\sim4000$ which corresponds to multipole $\ell\sim 300$\,\cite{Das:2020xke}. Neutrino self-interaction models fit the ACT EE data better in the $600<\ell<1000$ range which is relevant for SI mode interaction strength. Thus, it can be inferred from this analysis that ACT polarization data is the likely driver for the enhancement of the SI mode, which was pointed out in Ref.\,\cite{Kreisch:2022zxp}. For the ACT TT data, the preferred SI mode interaction strength, which yields a better fit for the multipole range $1600<\ell<2000$, corresponds to a smaller value of $\lgeff$ (see Fig. 2 in Ref.\,\cite{Kreisch:2022zxp}). 
The high multipole ACT data does not have large constraining power because of the large error bars as can be seen from the $\tilde{\chi}^2_{\ell,{\rm ACT}}$ plots (inset) in Fig.\,\ref{fig:residuals_tt} and \ref{fig:residuals_ee}. The increase in the significance of the SI mode is driven by the intermediate scales, rather than at smaller scales ($\ell>2500$) that were unexplored by Planck. 
\begin{figure}[t]
    \centering
    \includegraphics[width=0.49\linewidth]{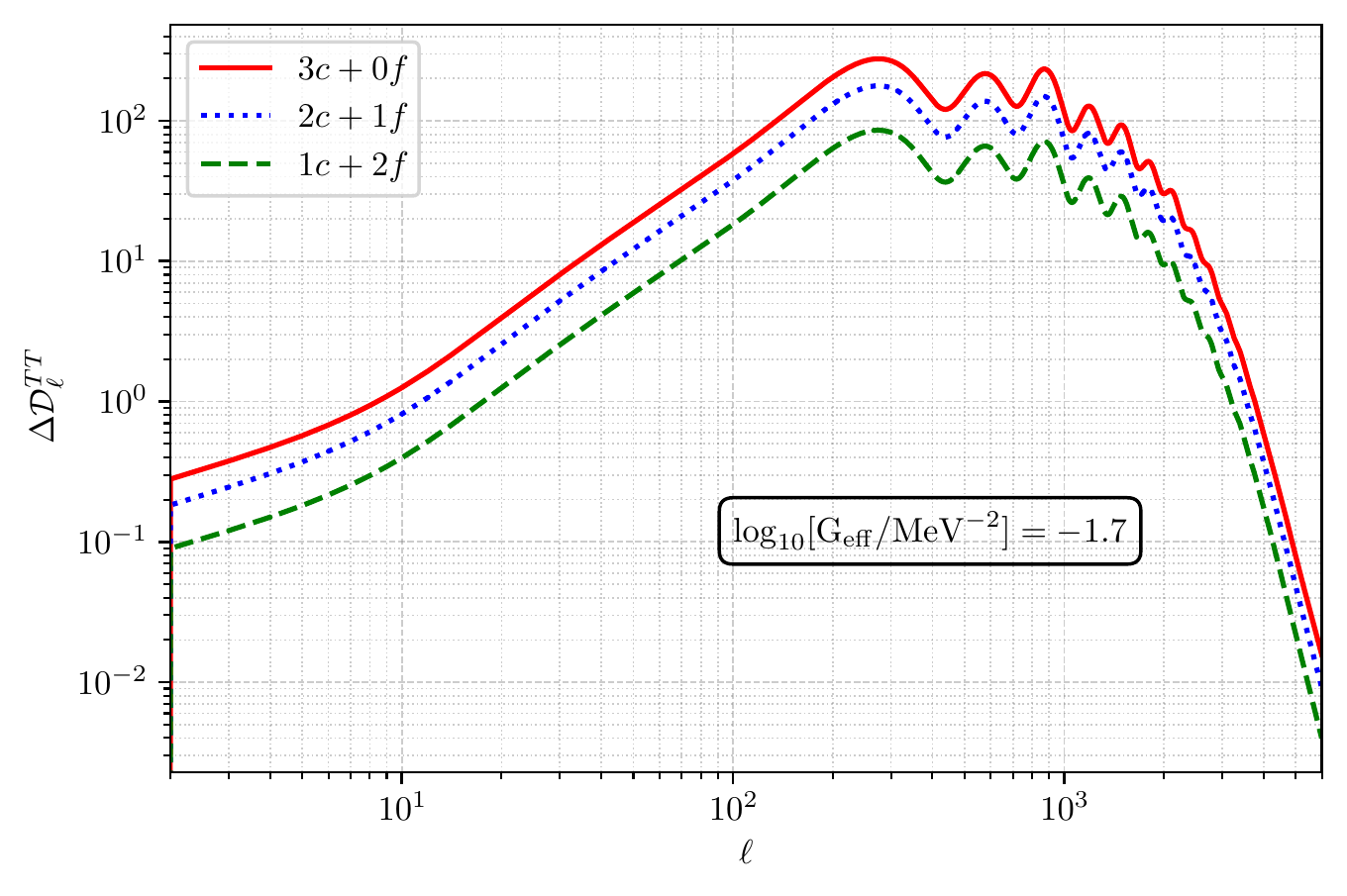}
    \includegraphics[width=0.49\linewidth]{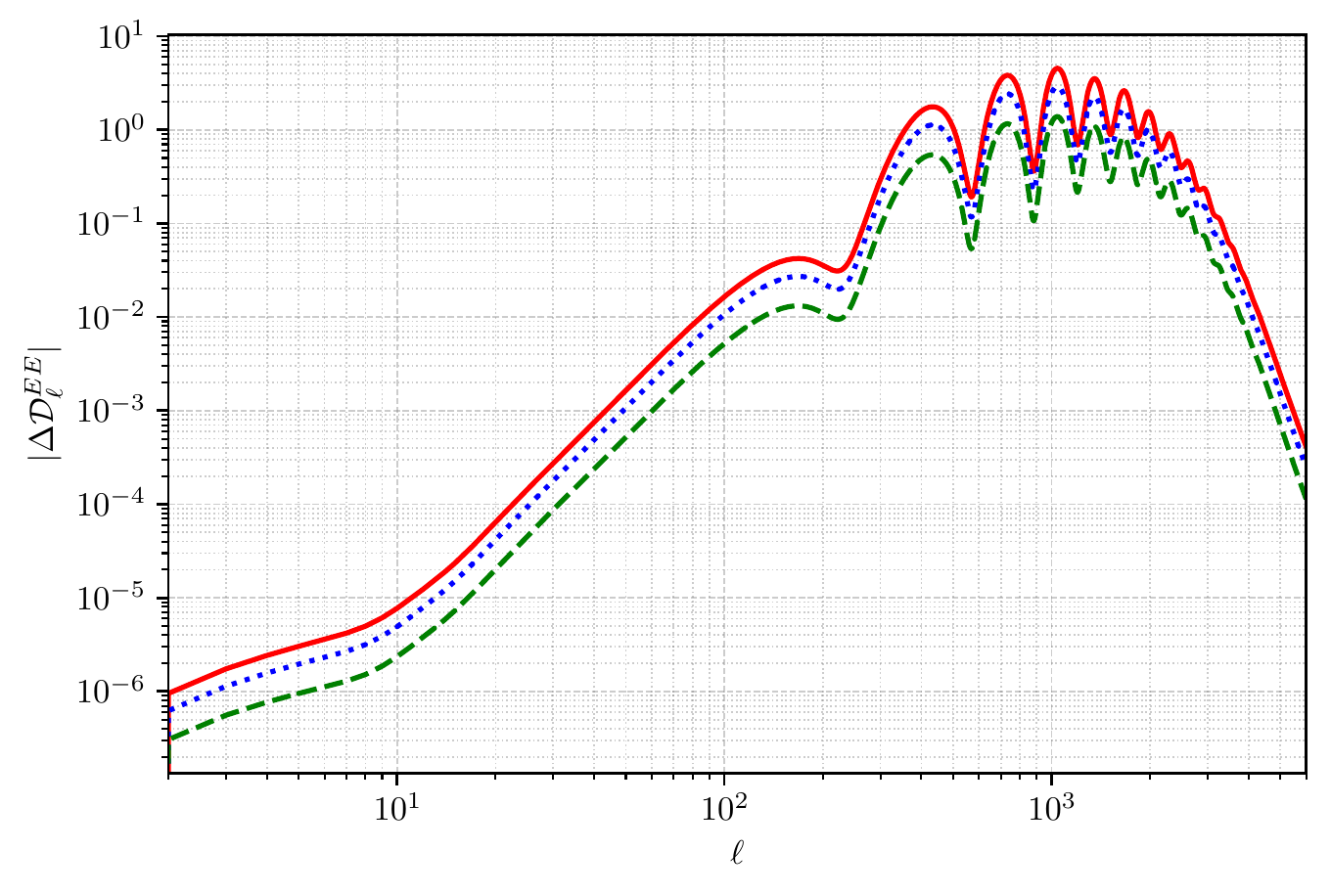}
    \caption{Changes in the TT and EE CMB power spectra $(D_\ell \equiv \ell(\ell+1)C_\ell/2\pi)$ due to neutrino self-interaction. The coupling strength is fixed at $\lgeff=-1$. Note the drop in $\Delta D_\ell$ above $\ell\gtrsim2000$.}
    \label{fig:abs_deltacl}
\end{figure}

At smaller scales, the CMB spectrum drops significantly due to diffusion damping. In Fig.\,\ref{fig:abs_deltacl}, we show the changes in the CMB spectra from \lcdm{} $(\Delta D_\ell)$ for flavor-specific scenarios. After $\ell\gtrsim 2000$, $\Delta D_\ell$ drops rapidly because the overall signal is damped. For example, the changes in the CMB spectrum due to SINU effects are $\Delta D_\ell \sim 10^{-2}$ at $\ell \sim 6000$. Probing such small changes at high multipole is going to be highly challenging. Future experiments will require high sensitivities and ultra-low angular resolution. In sec.~\ref{sec:future}, we will study the reach of a few future CMB measurements. Furthermore, for SI mode interaction strength, the neutrino decoupling takes place at a much smaller multipole. Therefore, more precise CMB measurement in the intermediate region $600<\ell<1000$ will be crucial to determine the fate of the SI mode.

The constraints derived in this section for the flavor-universal case are similar (approximately consistent within the error bar) to the ones derived in Ref.\,\cite{Kreisch:2022zxp} for the one-parameter extension. The slight discrepancies may be attributed to the different choice of prior ranges for $\lgeff$, the assumption on neutrino mass, the choice of $\tau_{reio}$-prior (specifically for the ACT-only analysis), and slight differences in dataset combination. For example, the ACT+Planck analysis in Ref.~\cite{Kreisch:2022zxp} does not include the Planck Lensing data, whereas this work does\footnote{In Ref.~\cite{Kreisch:2022zxp} Planck Lensing data was included in their analysis with the BAO data.}. 


\subsection{Effects of massive neutrino}
\begin{figure}[t]
    \centering
    \includegraphics[width=\textwidth]{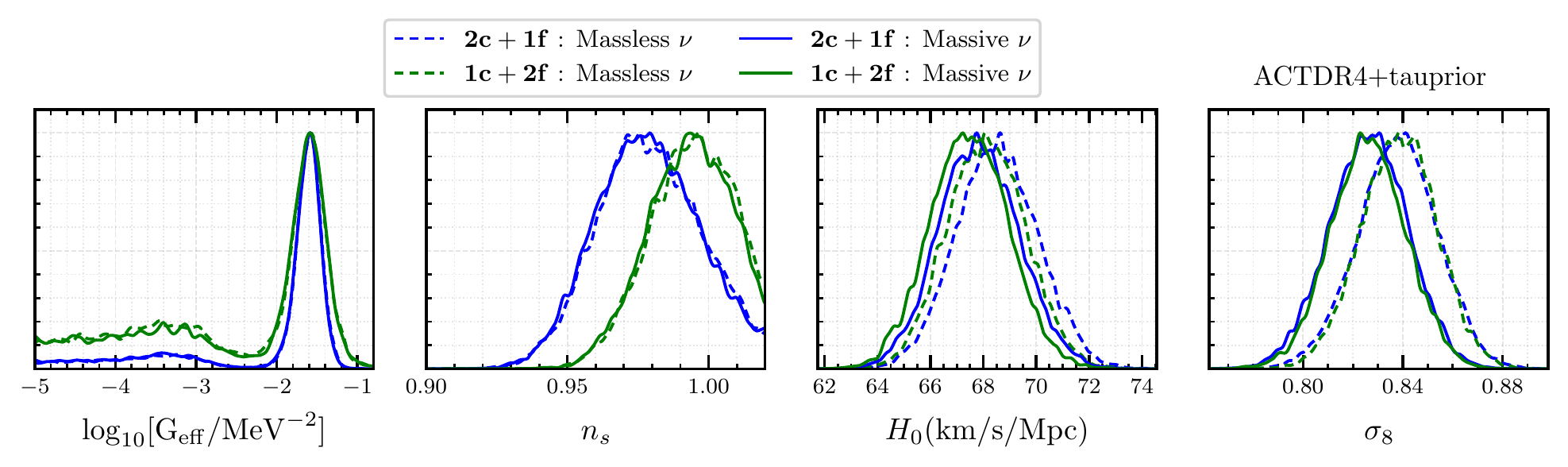}
    \includegraphics[width=\textwidth]{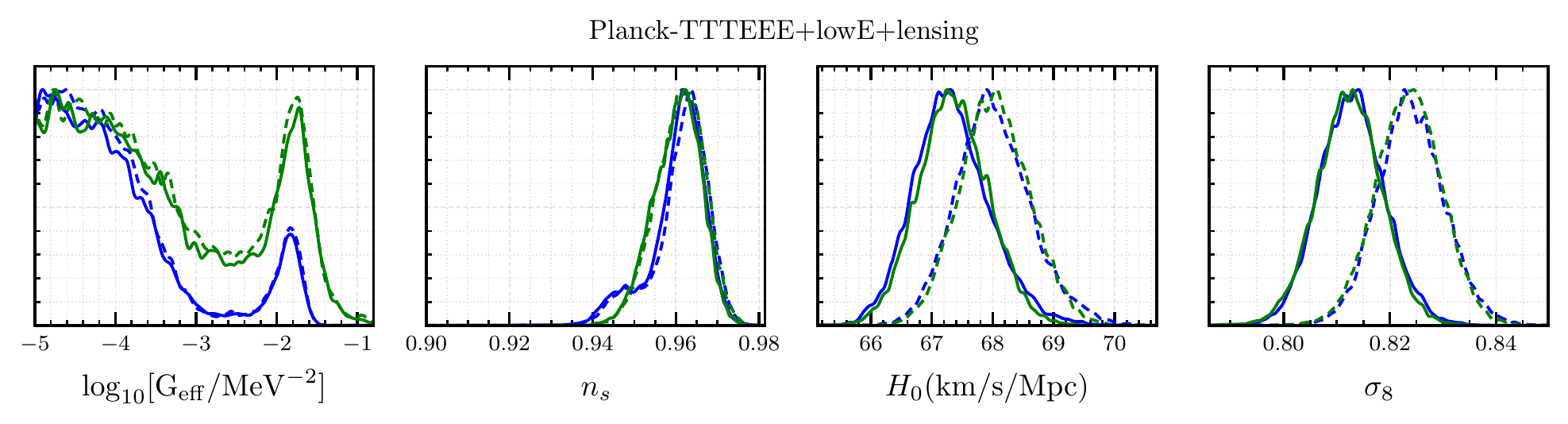}
    \includegraphics[width=\textwidth]{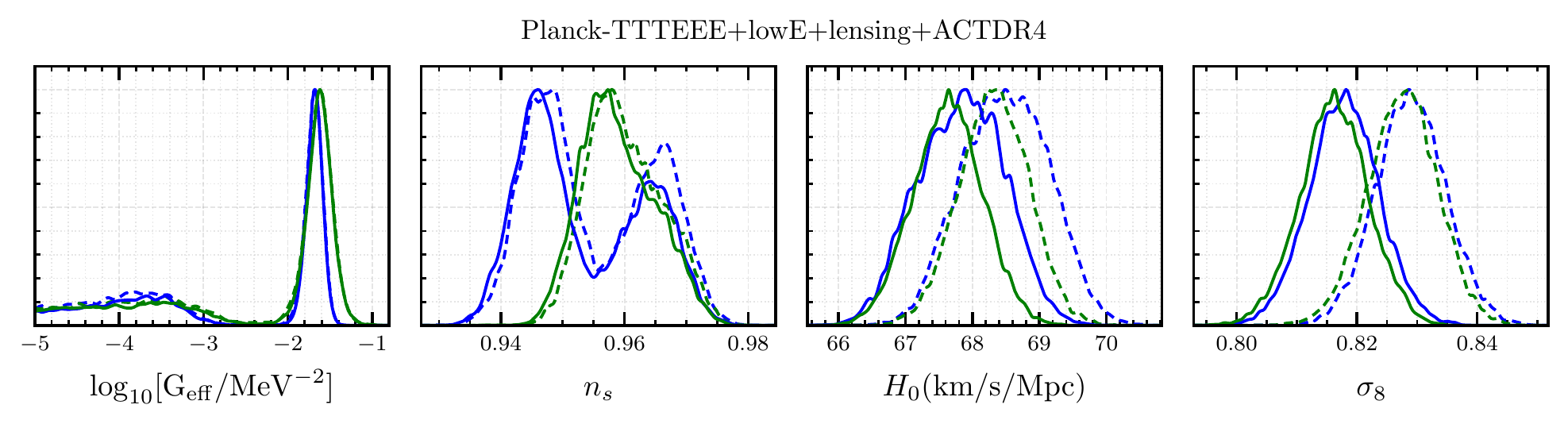}  
    \caption{Comparison between the posterior distributions from the analyses with massless (dashed) and massive (solid) neutrino in the \two{} (blue) and \one{} (green) cases. In massive case, for each of the models, one of the \emph{free streaming} neutrinos is modeled as massive with mass $0.06~{\rm eV}$. Adding neutrino mass does not change the $\lgeff$ posteriors by any discernible amount. See Fig.\,\ref{fig:triall_massive} in Appendix.~\ref{app:triplots} for an extended parameter plot for Planck + ACT analysis.}
    \label{fig:tri-act-planck-ml-ms}
\end{figure}
In this section, we show the results of a separate analysis including massive neutrino. For this purpose, we added the massive neutrino as a free-streaming neutrino with a fixed total mass, but not as coupled neutrinos.
Hence, we study the cosmology of \one{} and \two{} where one neutrino species among the respective $\mathbf{2f}$ and $\mathbf{1f}$ species is massive with a mass of $m_\nu = 0.06\,{\rm eV}$. 
We want to emphasize that such separation of the massive neutrinos and interacting neutrinos is not expected to give rise to any unusual results given the negligible correlation between $\lgeff$ and $\sum m_\nu$\,\cite{RoyChoudhury:2020dmd,Kreisch:2022zxp}.


Fig.\,\ref{fig:tri-act-planck-ml-ms} shows the 1D posterior of relevant parameters for \one{} and \two{} scenarios for all the datasets. From the $\lgeff$ posteriors, we see that adding neutrino mass does not change the favored interaction strength in SI and MI modes.
However, as expected with massive neutrinos, they increase the total matter density in the Universe and change the time of matter-radiation equality. This affects the background density evolution and results in slightly smaller values of $H_0$ and $\sigma_8$ relative to the massless case. The change in the matter-radiation equality also changes the size of the sound horizon at recombination and the distance to the last scattering surface. Together, these shift the CMB power spectra toward smaller $\ell$. The damping of the matter power spectrum at small scale additionally damps the CMB spectra at smaller angular scales\,\cite{Lesgourgues:2006nd}. Note that the shifts in $H_0$ and $\sigma_8$ due to neutrino mass are greater for the Planck dataset compared to ACT because of Planck's large-scale data.
In appendix~\ref{sec:mass_table}, we provide the constraints on all the cosmological parameters for the massive SINU scenario.


\subsection{Effects of varying $\neff$}
A well-motivated extension of the \lcdm{} cosmology is the addition of extra relativistic degrees of freedom, denoted by $\dneff$. The baseline model \lcdm{} has three active neutrinos which together yield $\neff=3.046$.

Because we have flavor-nonuniversal self-interaction, the additional neutrino states may or may not be self-interacting. In this section, we will study the effects of varying both the free streaming and coupled neutrino fraction in SINU scenarios. The effects of adding more radiation species are qualitatively opposite to that of neutrino self-interaction\,\cite{2013PhRvD..87h3008H,Kreisch:2019yzn,Das:2020xke}. Larger $\neff$ increases the net radiation density which results in a greater expansion rate of the Universe during radiation domination and a delay of the matter-radiation equality. This shifts the CMB power spectrum peaks toward low $\ell$ and damps the tail of the spectrum.

\subsubsection{Additional free streaming neutrinos}
\begin{figure}[t]
    \centering
    \includegraphics[width=\textwidth]{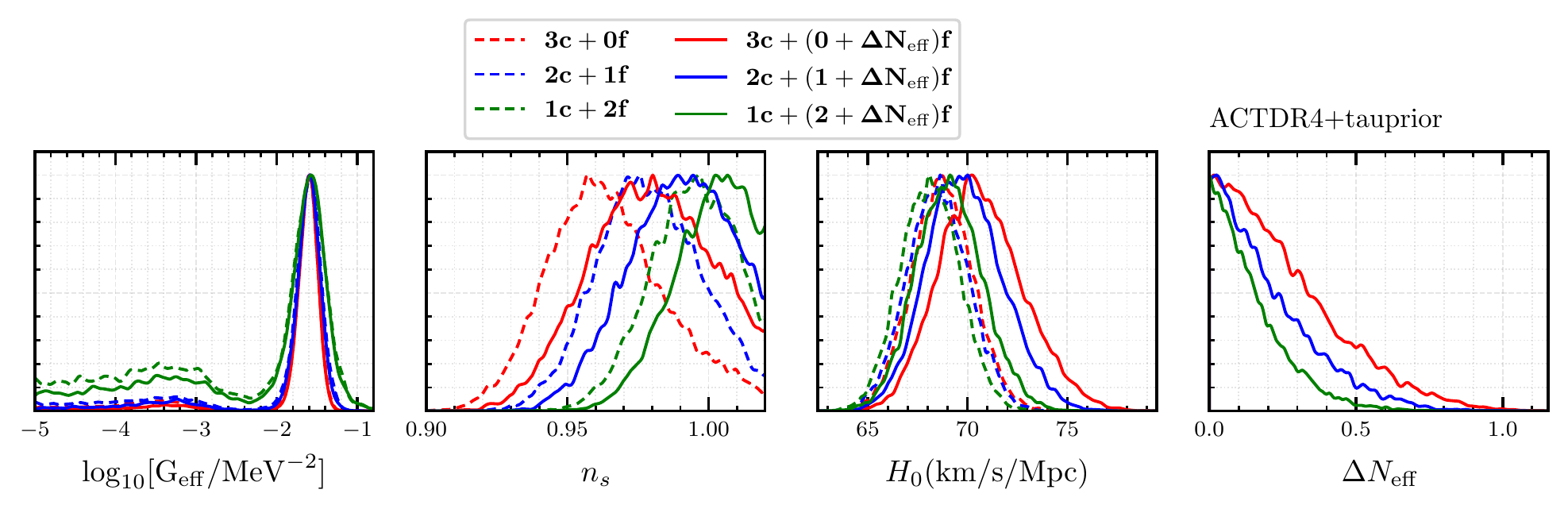}
    \includegraphics[width=\textwidth]{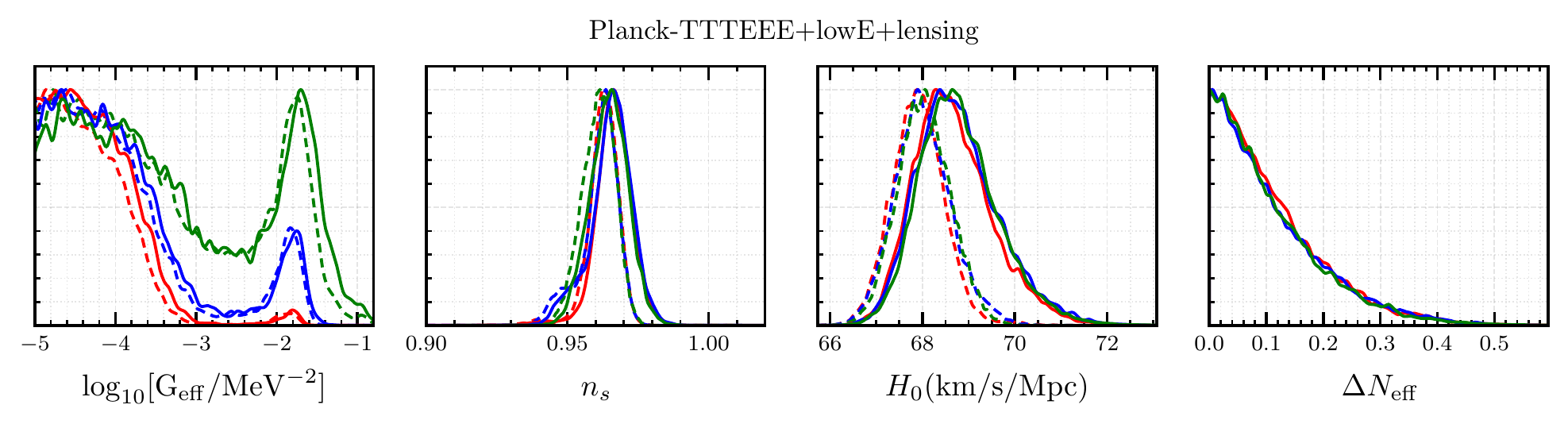}
    \includegraphics[width=\textwidth]{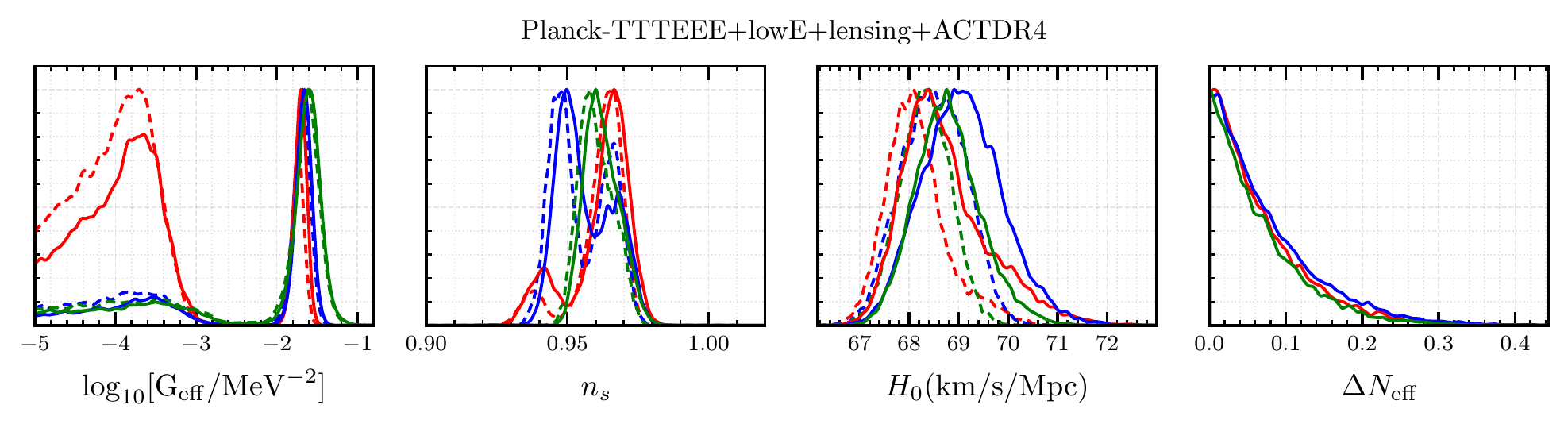}  
    \caption{Comparison between the posterior distributions from the analyses with fixed $\neff=3.046$ and varying $\neff$. The extra $\dneff$ component is treated as noninteracting.  See Fig.\,\ref{fig:triall_fsrad} in Appendix.~\ref{app:triplots} for an extended parameter plot for the varying $\neff$ analysis with Planck + ACT data. }
    \label{fig:tri-act-planck-ml-free}
\end{figure}
\begin{table}[h]
    \centering
    \resizebox{\textwidth}{!}{
    \begin{tabular}{|c|}
\hline
\\
\hline
Parameters \\
\hline
$10^2 \omega_{\rm b}$\\
$\omega_{\rm cdm}$\\
$100\theta{}_{s }$\\
$\ln10^{10}A_s$\\
$n_{s }$\\
$\tau_{\rm reio}$\\
$\log_{10} [{\rm G_{eff}} / {\rm MeV}^{-2}]$\\
$\Delta N_{\rm eff}$\\
\hline
$H_0 ({\rm km/s/Mpc})$\\
$\sigma_8$\\
\hline
$\chi^2 - \chi^2_{\overline{\Lambda{\rm CDM}}}$\\
\hline
$\Delta{\rm AIC}$\\
\hline
\end{tabular}
    \begin{tabular}{|P{2.6cm}P{2.6cm}P{2.6cm}|}
\hline
\multicolumn{3}{|c|}{ACTDR4+tauprior}\\
\hline
$3c+(0+\Delta N_{\rm eff})f$ & $2c+(1+\Delta N_{\rm eff})f$ & $1c+(2+\Delta N_{\rm eff})f$\\
\hline
$ 2.172\pm 0.034$ & $ 2.169\pm 0.033$ & $ 2.168\pm 0.031$\\
$ 0.1244^{+0.0044}_{-0.0052}$ & $ 0.1231^{+0.0043}_{-0.0048}$ & $ 0.1218^{+0.0037}_{-0.0045}$\\
$ 1.0477^{+0.0016}_{-0.00051}$ & $ 1.0459^{+0.0015}_{-0.00058}$ & $ 1.0442\pm 0.0011$\\
$ 2.989^{+0.024}_{-0.033}$ & $ 3.012^{+0.027}_{-0.030}$ & $ 3.035\pm 0.025$\\
$ 0.979^{+0.025}_{-0.022}$ & $ 0.989^{+0.022}_{-0.015}$ & $> 0.994$\\
$ 0.060\pm 0.010$ & $ 0.0599\pm 0.0098$ & $ 0.0601\pm 0.0096$\\
$ -1.79^{+0.32}_{+0.052}$ & $ -1.89^{+0.47}_{+0.12}$ & $ -2.3^{+1.1}_{-1.2}$\\
$< 0.315$ & $< 0.243$ & $< 0.172$\\
\hline
$ 70.7^{+1.9}_{-2.3}$ & $ 69.9^{+1.7}_{-2.1}$ & $ 69.1^{+1.5}_{-1.8}$\\
$ 0.849\pm 0.018$ & $ 0.846\pm 0.018$ & $ 0.845\pm 0.016$\\
\hline
$-4.61$ & $-3.54$ & $-0.49$\\
\hline
$-2.61$ & $-1.54$ & $1.51$\\
\hline
\end{tabular}
    \begin{tabular}{|P{2.6cm}P{2.6cm}P{2.6cm}|}
\hline
\multicolumn{3}{|c|}{Planck-TTTEEE+lowE+lensing}\\
\hline
$3c+(0+\Delta N_{\rm eff})f$ & $2c+(1+\Delta N_{\rm eff})f$ & $1c+(2+\Delta N_{\rm eff})f$\\
\hline
$ 2.244^{+0.015}_{-0.018}$ & $ 2.246^{+0.015}_{-0.019}$ & $ 2.246^{+0.015}_{-0.018}$\\
$ 0.1215^{+0.0014}_{-0.0020}$ & $ 0.1215^{+0.0013}_{-0.0021}$ & $ 0.1215^{+0.0014}_{-0.0020}$\\
$ 1.04174^{+0.00029}_{-0.00041}$ & $ 1.042001^{+0.000099}_{-0.00074}$ & $ 1.0422^{+0.0012}_{-0.0010}$\\
$ 3.045\pm 0.017$ & $ 3.042^{+0.021}_{-0.014}$ & $ 3.040\pm 0.018$\\
$ 0.9664\pm 0.0066$ & $ 0.9652^{+0.0083}_{-0.0054}$ & $ 0.9647^{+0.0070}_{-0.0064}$\\
$ 0.0544\pm 0.0074$ & $ 0.0544\pm 0.0074$ & $ 0.0545^{+0.0068}_{-0.0076}$\\
$< -4.06$ & $< -3.76$ & $< -2.63$\\
$< 0.133$ & $< 0.134$ & $< 0.130$\\
\hline
$ 68.68^{+0.64}_{-0.97}$ & $ 68.78^{+0.63}_{-1.1}$ & $ 68.80^{+0.64}_{-0.99}$\\
$ 0.8285^{+0.0065}_{-0.0078}$ & $ 0.8287^{+0.0068}_{-0.0077}$ & $ 0.8277^{+0.0067}_{-0.0078}$\\
\hline
$1.09$ & $1.17$ & $1.36$\\
\hline
$3.09$ & $3.17$ & $3.36$\\
\hline
\end{tabular}
    }
    \newline
    \vspace*{2ex}
    \newline
    \resizebox{0.58\textwidth}{!}{
    \begin{tabular}{|c|}
\hline
\\
\hline
Parameters \\
\hline
$10^2 \omega_{\rm b}$\\
$\omega_{\rm cdm}$\\
$100\theta{}_{s }$\\
$\ln10^{10}A_s$\\
$n_{s }$\\
$\tau_{\rm reio}$\\
$\log_{10} [{\rm G_{eff}} / {\rm MeV}^{-2}]$\\
$\Delta N_{\rm eff}$\\
\hline
$H_0 ({\rm km/s/Mpc})$\\
$\sigma_8$\\
\hline
$\chi^2 - \chi^2_{\overline{\Lambda{\rm CDM}}}$\\
\hline
$\Delta{\rm AIC}$\\
\hline
\end{tabular}
    \begin{tabular}{|P{2.6cm}P{2.6cm}P{2.6cm}|}
\hline
\multicolumn{3}{|c|}{Planck-TTTEEE+lowE+lensing+ACTDR4}\\
\hline
$3c+(0+\Delta N_{\rm eff})f$ & $2c+(1+\Delta N_{\rm eff})f$ & $1c+(2+\Delta N_{\rm eff})f$\\
\hline
$ 2.240^{+0.014}_{-0.016}$ & $ 2.243\pm 0.014$ & $ 2.241\pm 0.014$\\
$ 0.1210^{+0.0012}_{-0.0018}$ & $ 0.1215^{+0.0015}_{-0.0017}$ & $ 0.1210^{+0.0012}_{-0.0015}$\\
$ 1.04290^{-0.00049}_{-0.0013}$ & $ 1.0441^{+0.0018}_{-0.0024}$ & $ 1.04318^{+0.00099}_{-0.0014}$\\
$ 3.040^{+0.029}_{-0.0058}$ & $ 3.025\pm 0.025$ & $ 3.038^{+0.015}_{-0.018}$\\
$ 0.962^{+0.012}_{-0.0035}$ & $ 0.956^{+0.014}_{-0.012}$ & $ 0.9622^{+0.0054}_{-0.0072}$\\
$ 0.0552\pm 0.0058$ & $ 0.0549\pm 0.0058$ & $ 0.0552\pm 0.0058$\\
$ -3.58^{+2.0}_{-0.76}$ & $ -2.5^{+1.0}_{-1.3}$ & $ -2.29^{+0.99}_{-1.3}$\\
$< 0.0807$ & $< 0.0894$ & $< 0.0770$\\
\hline
$ 68.74^{+0.50}_{-1.1}$ & $ 69.09^{+0.74}_{-0.87}$ & $ 68.79^{+0.56}_{-0.74}$\\
$ 0.8328^{+0.0055}_{-0.0070}$ & $ 0.8318\pm 0.0064$ & $ 0.8299\pm 0.0062$\\
\hline
$3.71$ & $0.56$ & $0.4$\\
\hline
$5.71$ & $2.56$ & $2.4$\\
\hline
\end{tabular}
    }
        \caption{Mean values and 68\% confidence limits for SINU with additional free streaming neutrinos. Note that the values and limits are for the full sample where we have not separated out the MI and SI modes. We also show the $\chi^2$ difference from the bestfit 7-parameter $\Lambda$CDM model where $N_{\rm eff}$ is also varied (denoted as $\overline{\Lambda \rm CDM}$). The corresponding $\Delta$AIC is also shown.}\label{tab:fsrad}
\end{table}
In this section, we added extra free-streaming radiation $(\Delta N_{\rm eff})$ in all SINU scenarios and dubbed them as: $\mathbf{3c+(0+\Delta N_{\rm eff})f}$, $\mathbf{2c+(1+\Delta N_{\rm eff})f}$ and $\mathbf{1c+(2+\Delta N_{\rm eff})f}\,$\footnote{We have added a flat prior on $\Delta N_{\rm eff}$ for all SINU scenarios in the range $[0,1]$ }.
The results of adding extra free streaming are fairly simple to understand as it just increases $\neff$ without changing the amount of self-interaction. In Fig.\,\ref{fig:tri-act-planck-ml-free}, we show the 1D posteriors of all parameters using the three dataset combinations. The most significant changes happen in $n_s$ and $H_0$. The extra free streaming $\dneff$ increases their values, which is expected from the change in the matter-radiation equality and the sound horizon during recombination. Even though the ACT data prefer the SI mode and $\dneff$ cancels the effects of $\geff$, we do not see any increase in the significance or the position of the SI mode in $\geff$ in Fig.\,\ref{fig:tri-act-planck-ml-free}. This suggests that the effects of $\Delta N_{\rm eff}$ are compensated mostly by $n_s$.  In the \three{} case, there is a preference for larger $\dneff$ than the other cases and, as a result, larger $H_0$ too.

When only Planck data is used, the difference between the three SINU cases goes away almost completely for all parameters except for $\lgeff$. This is because the dominant modes, the MI mode of \three{} and the SI mode of \one{}, are close to each other for other parameters. Finally, in none of the cases shown in Fig.\,\ref{fig:tri-act-planck-ml-free}, we find any preference for a nonzero value of $\dneff$.

\subsubsection{Additional self-interacting neutrinos}

As in our setup, we always use an integer number of self-interacting neutrinos to assign their individual coupling strengths, we cannot directly use the $\neff$ parameter to vary their number. However, as far as the total relativistic energy density of free streaming species is concerned, it can be varied either by changing the number of states or the temperatures. So, we use the relative temperature $\xi_i\equiv T_i/T_\gamma$ to vary the energy density of the \emph{coupled} neutrinos\footnote{The (flat) prior range for the temperature of the coupled was chosen as $\xi_{\nu,c} = [0.0,1.0]$. Only for this subsection, we varied $H_0$ instead of $100\theta_s$ with a flat prior between $[58, 78]$.}. This is possible only for a thermal species. In fact, from a particle physics perspective, varying $\xi_i$ is more physical as the number of flavor eigenstates cannot be non-integer.

However, when a radiation species is self-interacting with a temperature-dependent interaction rate, varying $\xi_i$ is not straightforward. For free streaming neutrinos, there is a complete degeneracy between temperature and the number of states since the relevant cosmological quantities like energy density, pressure density depends on the combination $N_\nu T_\nu^4$, where $N_\nu$ is the number of free-streaming neutrino states. On the other hand, for self-interacting neutrinos the interaction rate has a different dependence of $T_\nu$ as shown in Eq.(\ref{eq:taudot}):
\begin{equation}
    \dot{\tau}_\nu = -a(\geff)^2T_\nu^5\;.
\end{equation}
Therefore, when $T_\nu$ (and hence $\xi_\nu$) is varied, to achieve the same interaction rate, $\geff$ also needs to be scaled according to Eq.(\ref{eq:taudot}). In other words, since $\dot{\tau}_\nu$ is the physical quantity, the cosmological signatures depend on the combination $(\geff)^2T_\nu^5$. 
So, to compare with the fixed temperature baseline model, we define a rescaled coupling strength
\begin{equation}
    \label{eq:pseudogeff}
    \gefft \equiv \geff\left(\frac{T_\nu}{T_{\nu,{\rm SM}}}\right)^{5/2}\;,
\end{equation}
\begin{figure}[t]
    \centering
    \includegraphics[width=\textwidth]{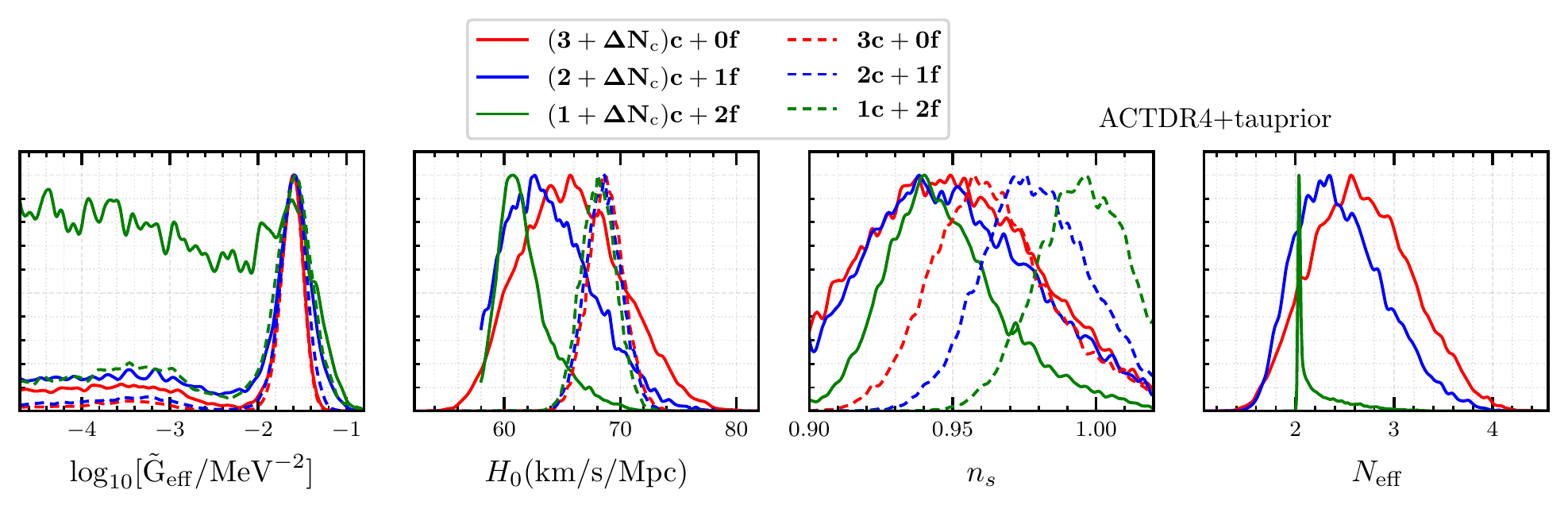}
    \includegraphics[width=\textwidth]{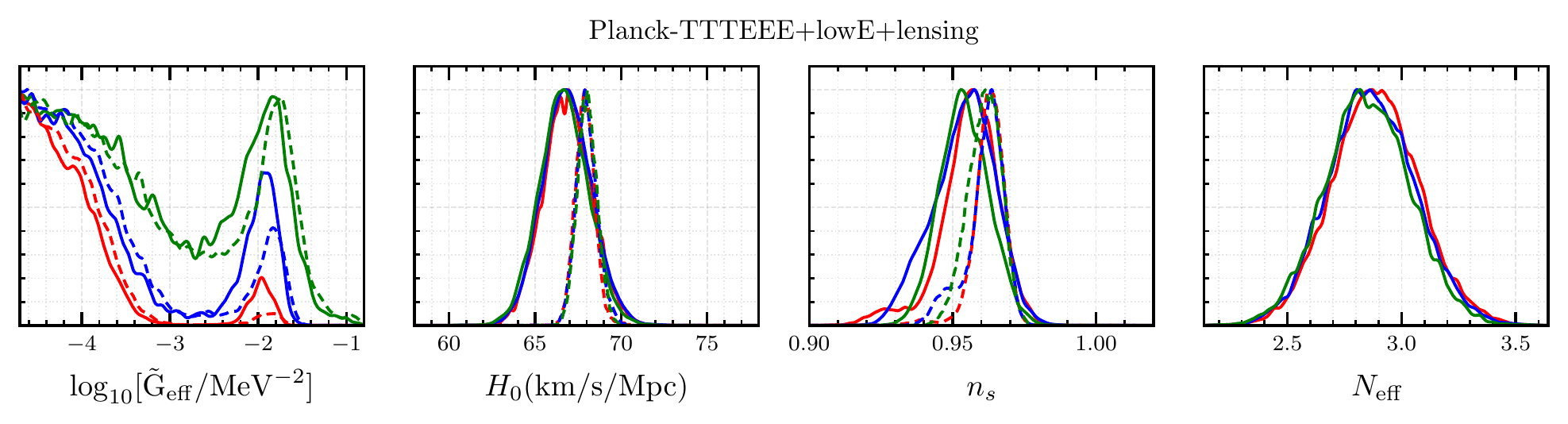}
    \includegraphics[width=\textwidth]{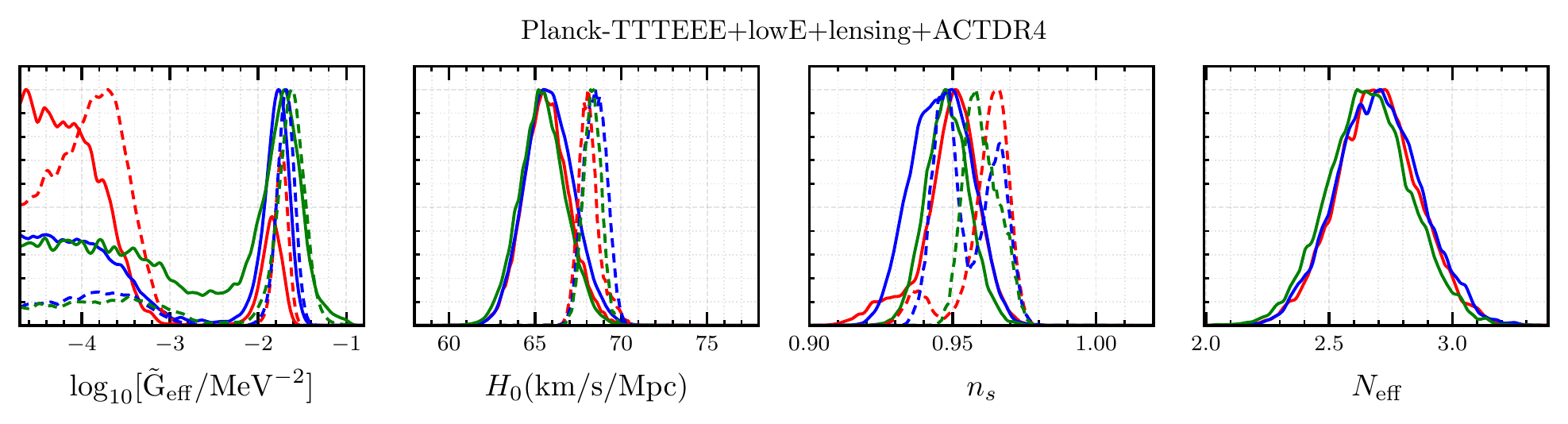}  
    \caption{The posterior distributions when the $\dneff$ for coupled neutrinos is varied. For comparison, we also show the respective distributions for fixed $\neff=3.046$. See Fig.\,\ref{fig:triall_nfsrad} in Appendix.~\ref{app:triplots} for an extended parameter plot for the varying $\neff$ analysis with Planck + ACT data.}
    \label{fig:1d_nfs}
\end{figure}where $T_{\nu,{\rm SM}}$ is the SM neutrino temperature. Thus, for the baseline base, $\gefft = \geff$. In Appendix.~\ref{app:gefft}, we explicitly demonstrate via MCMC analysis the degeneracy between $\geff$ and $T_\nu$. As an example, we studied the \three{} scenario with varying temperatures via two parametrizations: $N_c = 1$ and $N_c = 3$ coupled neutrinos with their corresponding varying temperature. We show that after proper rescaling (as shown in Eq.(\ref{eq:pseudogeff})), both models give the exact same constraints on \emph{all} cosmological parameters (Fig.\,\ref{fig:temp_comp_planck}).
\begin{table}
    \centering
    \resizebox{\textwidth}{!}{
    \begin{tabular}{|c|}
\hline
\\
\hline
Parameters \\
\hline
$10^2 \omega_{\rm b}$\\
$\omega_{\rm cdm}$\\
$100\theta{}_{s }$\\
$\ln10^{10}A_s$\\
$n_{s }$\\
$\tau_{\rm reio}$\\
$\log_{10} [{\rm G_{eff}} / {\rm MeV}^{-2}]$\\
$\frac{T_{{\rm c},1\nu}} {T_{{\rm f},1\nu}}$\\
\hline
$\log_{10} [{\rm \tilde{G}_{eff}} / {\rm MeV}^{-2}]$\\
$H_0 ({\rm km/s/Mpc})$\\
$\sigma_8$\\
$\Delta N_{\rm c}$\\
\hline
$\chi^2 - \chi^2_{\overline{\Lambda{\rm CDM}}}$\\
\hline
$\Delta{\rm AIC}$\\
\hline
\end{tabular}
    \begin{tabular}{|P{2.6cm}P{2.6cm}P{2.6cm}|}
\hline
\multicolumn{3}{|c|}{ACTDR4+tauprior}\\
\hline
$(3 +\Delta N_{\rm c})c+0f$ & $(2 +\Delta N_{\rm c})c+1f$ & $(1 +\Delta N_{\rm c})c+2f$\\
\hline
$ 2.123\pm 0.051$ & $ 2.109^{+0.043}_{-0.051}$ & $ 2.089^{+0.030}_{-0.039}$\\
$ 0.1157^{+0.0079}_{-0.0091}$ & $ 0.1122^{+0.0062}_{-0.0083}$ & $ 0.1084^{+0.0032}_{-0.0052}$\\
$ 1.0472\pm 0.0021$ & $ 1.0462^{+0.0017}_{-0.0011}$ & $ 1.04516^{+0.00086}_{-0.00070}$\\
$ 2.987^{+0.030}_{-0.038}$ & $ 2.999^{+0.027}_{-0.030}$ & $ 3.012\pm 0.024$\\
$ 0.951^{+0.025}_{-0.034}$ & $ 0.951^{+0.025}_{-0.034}$ & $ 0.948^{+0.014}_{-0.025}$\\
$ 0.0583^{+0.010}_{-0.0093}$ & $ 0.0582\pm 0.0097$ & $ 0.0578\pm 0.0096$\\
$ -2.5^{+1.2}_{-1.7}$ & $ -2.5^{+1.4}_{-2.1}$ & $ -2.8\pm 1.2$\\
$ 0.966^{+0.049}_{-0.045}$ & $ 0.913\pm 0.070$ & $< 0.658$\\
\hline
$ -2.5^{+1.2}_{-1.4}$ & $ -2.6^{+1.4}_{-1.8}$ & $ -4.0^{+2.5}_{-1.4}$\\
$ 65.8^{+3.8}_{-5.0}$ & $ 64.1^{+2.4}_{-4.7}$ & $ 61.9^{+1.1}_{-2.7}$\\
$ 0.825\pm 0.026$ & $ 0.814^{+0.022}_{-0.026}$ & $ 0.803^{+0.015}_{-0.019}$\\
$ -0.36\pm 0.50$ & $ -0.57^{+0.31}_{-0.57}$ & $ -0.830^{+0.032}_{-0.20}$\\
\hline
$-4.73$ & $-3.76$ & $-1.07$\\
\hline
$-2.73$ & $-1.76$ & $0.93$\\
\hline
\end{tabular}
    \begin{tabular}{|P{2.6cm}P{2.6cm}P{2.6cm}|}
\hline
\multicolumn{3}{|c|}{Planck-TTTEEE+lowE+lensing}\\
\hline
$(3 +\Delta N_{\rm c})c+0f$ & $(2 +\Delta N_{\rm c})c+1f$ & $(1 +\Delta N_{\rm c})c+2f$\\
\hline
$ 2.223\pm 0.022$ & $ 2.223\pm 0.022$ & $ 2.220\pm 0.022$\\
$ 0.1177\pm 0.0029$ & $ 0.1175\pm 0.0029$ & $ 0.1172\pm 0.0029$\\
$ 1.04246^{+0.00028}_{-0.00081}$ & $ 1.04275^{+0.00019}_{-0.0012}$ & $ 1.04266^{+0.00071}_{-0.00090}$\\
$ 3.030^{+0.021}_{-0.015}$ & $ 3.026^{+0.025}_{-0.018}$ & $ 3.027\pm 0.018$\\
$ 0.956^{+0.011}_{-0.0076}$ & $ 0.954^{+0.012}_{-0.0091}$ & $ 0.9542^{+0.0081}_{-0.0091}$\\
$ 0.0526\pm 0.0074$ & $ 0.0526\pm 0.0074$ & $ 0.0525\pm 0.0073$\\
$ -4.25^{+0.20}_{-0.73}$ & $ -3.8^{+2.0}_{-1.1}$ & $ -3.4^{+1.5}_{-1.2}$\\
$ 0.987^{+0.017}_{-0.015}$ & $ 0.977^{+0.027}_{-0.023}$ & $ 0.944^{+0.063}_{-0.044}$\\
\hline
$ -4.26^{+0.18}_{-0.73}$ & $ -3.9^{+2.0}_{-1.2}$ & $ -3.4^{+1.8}_{-1.6}$\\
$ 66.9\pm 1.4$ & $ 66.9\pm 1.4$ & $ 66.7\pm 1.4$\\
$ 0.8171^{+0.0093}_{-0.011}$ & $ 0.8168^{+0.0093}_{-0.011}$ & $ 0.815\pm 0.010$\\
$ -0.16\pm 0.19$ & $ -0.18\pm 0.19$ & $ -0.19\pm 0.19$\\
\hline
$0.25$ & $0.19$ & $0.15$\\
\hline
$2.25$ & $2.19$ & $2.15$\\
\hline
\end{tabular}
    }
    \newline
    \vspace*{2ex}
    \newline
    \resizebox{0.58\textwidth}{!}{
    \begin{tabular}{|c|}
\hline
\\
\hline
Parameters \\
\hline
$10^2 \omega_{\rm b}$\\
$\omega_{\rm cdm}$\\
$100\theta{}_{s }$\\
$\ln10^{10}A_s$\\
$n_{s }$\\
$\tau_{\rm reio}$\\
$\log_{10} [{\rm G_{eff}} / {\rm MeV}^{-2}]$\\
$\frac{T_{{\rm c},1\nu}} {T_{{\rm f},1\nu}}$\\
\hline
$\log_{10} [{\rm \tilde{G}_{eff}} / {\rm MeV}^{-2}]$\\
$H_0 ({\rm km/s/Mpc})$\\
$\sigma_8$\\
$\Delta N_{\rm c}$\\
\hline
$\chi^2 - \chi^2_{\overline{\Lambda{\rm CDM}}}$\\
\hline
$\Delta{\rm AIC}$\\
\hline
\end{tabular}
    \begin{tabular}{|P{2.6cm}P{2.6cm}P{2.6cm}|}
\hline
\multicolumn{3}{|c|}{Planck-TTTEEE+lowE+lensing+ACTDR4}\\
\hline
$(3 +\Delta N_{\rm c})c+0f$ & $(2 +\Delta N_{\rm c})c+1f$ & $(1 +\Delta N_{\rm c})c+2f$\\
\hline
$ 2.202\pm 0.020$ & $ 2.204\pm 0.021$ & $ 2.200\pm 0.020$\\
$ 0.1150\pm 0.0026$ & $ 0.1152^{+0.0025}_{-0.0029}$ & $ 0.1145\pm 0.0026$\\
$ 1.04326^{+0.00010}_{-0.00096}$ & $ 1.0439^{+0.0021}_{-0.0016}$ & $ 1.04338^{+0.00074}_{-0.00063}$\\
$ 3.026^{+0.022}_{-0.011}$ & $ 3.019^{+0.026}_{-0.022}$ & $ 3.025^{+0.014}_{-0.016}$\\
$ 0.949^{+0.011}_{-0.0066}$ & $ 0.946\pm 0.010$ & $ 0.9483^{+0.0070}_{-0.0079}$\\
$ 0.0533\pm 0.0058$ & $ 0.0532\pm 0.0057$ & $ 0.0531\pm 0.0057$\\
$ -4.06^{+0.15}_{-0.89}$ & $ -3.3\pm 1.2$ & $ -2.9^{+1.6}_{-2.0}$\\
$ 0.970\pm 0.015$ & $ 0.954\pm 0.024$ & $ 0.882^{+0.070}_{-0.048}$\\
\hline
$ -4.09^{+0.15}_{-0.90}$ & $ -3.3\pm 1.2$ & $ -3.1^{+1.6}_{-2.0}$\\
$ 65.7^{+1.2}_{-1.4}$ & $ 65.9\pm 1.4$ & $ 65.4\pm 1.3$\\
$ 0.8139\pm 0.0094$ & $ 0.8144\pm 0.0095$ & $ 0.8113\pm 0.0093$\\
$ -0.35\pm 0.17$ & $ -0.34\pm 0.17$ & $ -0.38^{+0.16}_{-0.18}$\\
\hline
$-0.06$ & $-2.08$ & $-2.26$\\
\hline
$1.94$ & $-0.08$ & $-0.26$\\
\hline
\end{tabular}
    }
        \caption{Mean values and 68\% confidence limits for SINU with additional self-interacting neutrinos. Note that the values and limits are for the full sample where we have not separated out the MI and SI modes. In addition to the six \lcdm{} parameters here we have varied $\lgeff$ and $\xi_\nu$ for the coupled neutrino flavor. Here $\Delta N_c \equiv N_{\rm eff} - 3.046 = \Delta N_{\rm eff}$. We also show the $\chi^2$ difference from the bestfit 7-parameter $\Lambda$CDM model where $N_{\rm eff}$ is also varied (denoted as $\overline{\Lambda \rm CDM}$). The corresponding $\Delta$AIC is also shown. ${T_{{\rm c},1\nu}} / {T_{{\rm f},1\nu}}$ in the table is the ratio of coupled to the free streaming temperature (SM neutrino temperature) and is proportional to $\xi_\nu$ for the coupled flavor. For all datasets ${T_{{\rm c},1\nu}} / {T_{{\rm f},1\nu}} < 1$ (equivalently $\Delta N_c<0$) and thus the mean values of $\log_{10} [{\rm \tilde{G}_{eff}} / {\rm MeV}^{-2}]$ are always lower than the corresponding $\log_{10} [{\rm G_{eff}} / {\rm MeV}^{-2}]$ following Eq.~\eqref{eq:pseudogeff}. }\label{tab:nfsrad}
\end{table}

This discussion is also relevant for comparison with existing results in the literature for the varying $\neff$ (coupled) scenarios for flavor universal coupling. Since those analyses usually do not vary the neutrino temperature, but rather the number of states directly, the $\geff$ does not scale in those cases. Therefore, $\geff$ in those analyses (where $\neff$ is varied directly), should be compared with $\gefft$ in our analysis. Note that, for a \emph{physical} model where $\neff$ is varied through the temperature of the species, the scaling in Eq.(\ref{eq:pseudogeff}) should be properly taken into account. For the analysis in this section, we have studied the temperature-varying coupled neutrinos as $\mathbf{(3 +\Delta N_{\rm c})c+0f}$, $\mathbf{(2 +\Delta N_{\rm c})c+1f}$ and  $\mathbf{(1 +\Delta N_{\rm c})c+2f}$ cases. In each of the scenarios, only the temperatures of the corresponding \emph{coupled} species were varied.

In Fig.\,\ref{fig:1d_nfs} we show the constraints on some relevant cosmological parameters along with $\lgefft$. It has been shown that the ACT data prefers a smaller value of $N_{\rm eff}$ in general\,\cite{ACT:2020gnv}. This also can be seen from the $\neff$ plots. This happens due to the strong degeneracy with $n_s$ which arise from the lack of large-scale structure data. In general, Planck and ACT both prefer a smaller value of $\neff$ compared to the standard value of 3.046. Therefore, due to the less amount of silk damping the preferred value of $n_s$ are smaller compared to the baseline models. The decrease in $\neff$ also results in a smaller value of the $H_0$ worsening the $H_0$ tension in this case. The central values for SI mode in $\geff$ moved slightly to the smaller values and increased in significance for all datasets. The shift in SI mode $\geff$ can be understood through the additional degeneracy with $\neff$. In this scenario, the changes induced by $n_s$ are compensated by both $\neff$ and $\geff$ (or $\gefft$). Thus, the additional freedom of varying $\neff$ limits the values allowed for $\gefft$ and pushes the peak to a slightly smaller value. Note that this freedom is also present in the free-streaming $\dneff$ result in the earlier section. However, in that case, because only positive $\dneff $ is allowed, the role of $\dneff $ to compensate for the changes in the CMB spectrum is quite limited. Therefore, the SI mode values of $\geff$ are virtually unmodified from the baseline model in that case. This constrained fit is also reflected in the $\chi^2$ difference from \lcdm{} in Table\,\ref{tab:fsrad} and \ref{tab:nfsrad} where we see that coupled $\neff$ gives a better fit of the data compared to free-streaming $\neff$ for each of the respective scenarios.

\section{Prospects of future CMB experiments}
\label{sec:future}
As demonstrated in the previous section, the main constraining power of current CMB experiments on the neutrino self-interaction comes from the intermediate $500\lesssim\ell\lesssim 1500$ range of the multipoles. Improving the constraint on $\geff$ would require more precise measurements of the CMB spectra, especially the EE spectrum, to determine the amplitude and the position of the peaks more accurately. Even though neutrino self-interaction is not a part of the primary science goals of any of the upcoming CMB experiments, instrument requirements, and survey strategies for other goals such as primordial gravitational wave and light relics could benefit the search for signatures of SINU. CMB-S4 plans to constrain $\dneff < 0.06$ at 95\% confidence level by conducting a high-resolution survey, and will also perform an ultra-low noise survey aimed for primordial gravitational wave\,\cite{Abazajian:2019eic}. Both CMB-S4 and CORE are aiming for $\sim\mathcal{O}$(few) $\muKam$ sensitivity in the mid-frequency range. Such strategies, if realized, will certainly improve the constraint on $\geff$. In this section, we use the planned design specifications of CMB-S4 and CORE to forecast future constraints on $\geff$. We also show the limit for a cosmic-variance limited experiment.


\begin{figure}[t]
    \centering
    \includegraphics[width=0.6\textwidth]{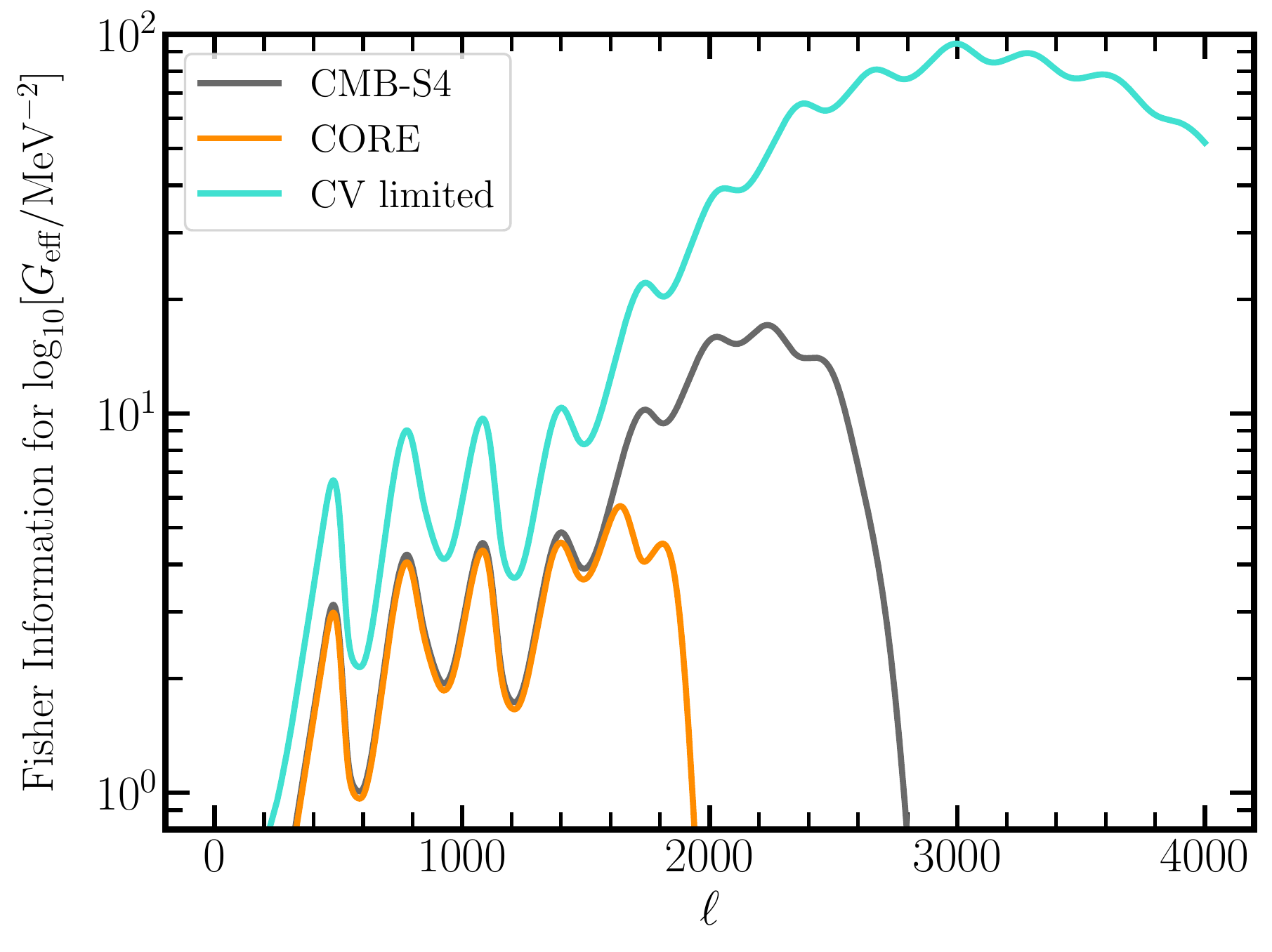}
    \caption{The distribution of Fisher information for $\lgeff$ over the multipoles $\ell$ for CMB-S4, CORE, and cosmic variance limited experiments. The instrumental noise dominates over cosmic variance for $\ell\gtrsim2000$ which shows that a large portion of the information from the high $\ell$ tail of the CMB anisotropy power spectra would still be unavailable in upcoming experiments.}
    \label{fig:fisher_info_geff}
\end{figure}

\subsection{Fisher forecast}
We use the Fisher formalism to forecast limits using the same set of seven parameters used in the previous section: $\{H_0, \omb, \omc, A_s, n_s, \treio, \lgeff\}$. The Fisher matrix can be written as\,\cite{Kamionkowski:1996ks,Li:2018zdm}
\begin{equation}\label{eq:fisher}
    F_{ij} = \sum_\ell \dfrac{2\ell+1}{2}~f_\mathrm{sky}~\mathrm{Tr}\left(\bc^{-1}\dfrac{\partial\bc}{\partial\lambda_i} \bc^{-1}\dfrac{\partial\bc}{\partial\lambda_j}\right)\,,
\end{equation}
where $i,j$ are the parameter indices, $\lambda_i$ are the parameters, $f_\mathrm{sky}$ is the fraction of the sky observed by the experiment. The covariance matrix $\bc$ is defined as
\begin{equation}
    \bc =
    \begin{bmatrix}
        C_\ell^{TT} & C_\ell^{TE} \\
        C_\ell^{TE} & C_\ell^{EE} 
    \end{bmatrix}
    + \mathbf{N}_\ell\,,
\end{equation}
with a diagonal noise covariance 
matrix $\mathbf{N}_\ell = \mathrm{diag}(N_\ell^{TT}, N_\ell^{EE})$. We do not include the lensing data. The noise can be taken to be\,\cite{Kamionkowski:1996ks}
\begin{equation}
    N_\ell^{TT} = \Delta_T^2 \exp\left(\ell(\ell+1)\dfrac{\theta_\mathrm{FWHM}^2}{8\ln2}\right)\,,
\end{equation}
where $\Delta_T$ is the instrumental noise in $\mu$K-radian and $\theta_\mathrm{FWHM}$ is the beam width. The polarization noise is $\Delta_P = \sqrt{2}\Delta_T$. CMB experiments observe in multiple frequency channels with the corresponding beam size and sensitivity depending on the frequency. We use the publicly available design parameters for CMB-S4 and CORE\,\cite{s4-wiki,CORE:2017oje}.
\begin{figure}[t]
    \centering
    \includegraphics[width=0.8\textwidth]{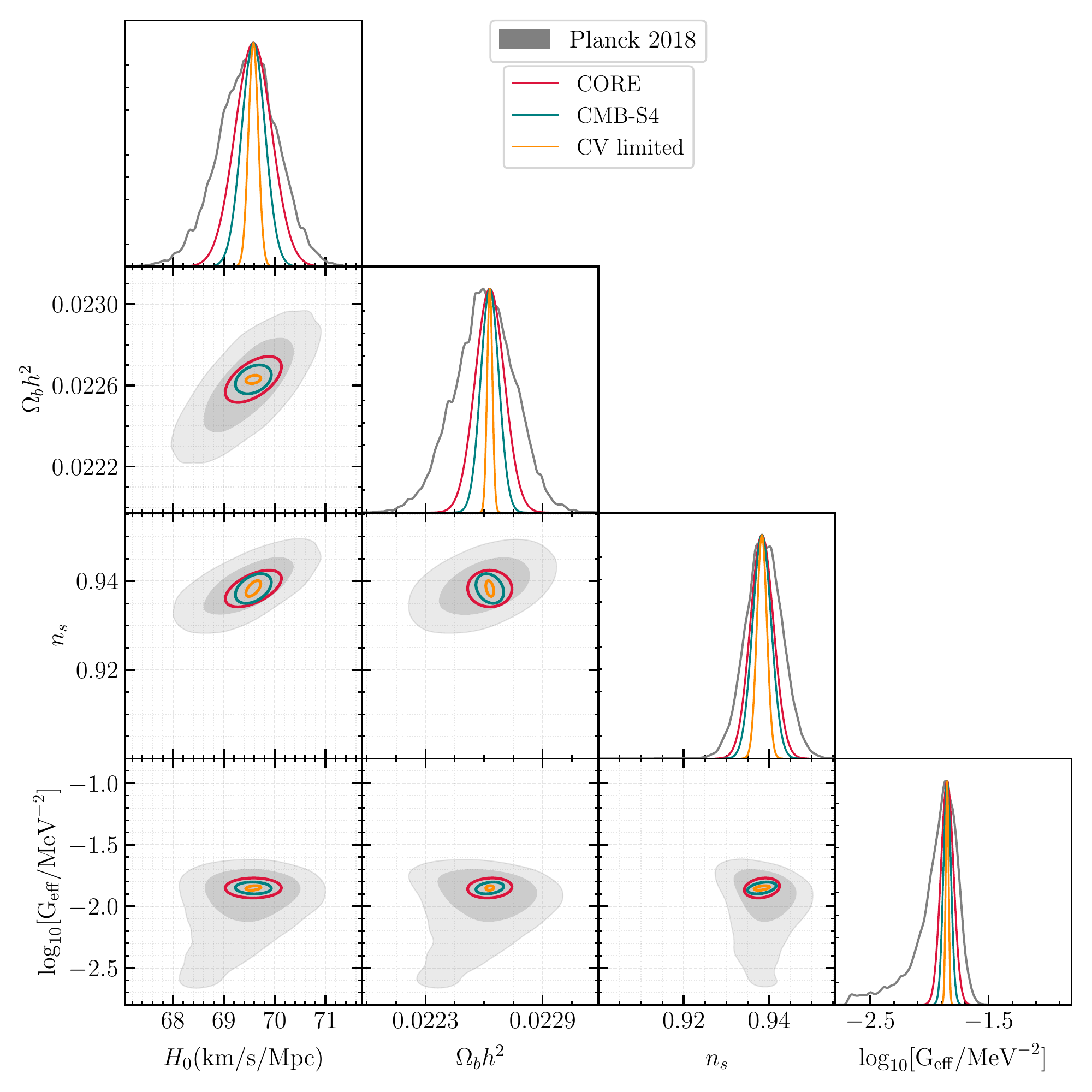}
    \caption{Fisher forecast for a few cosmological parameters for CMB-S4 (green), CORE (red), and cosmic variance limited (orange) experiments. Planck 2018 TTTEEE+lensing posteriors are shown in gray for comparison.}
    \label{fig:fisher_corner}
\end{figure}
Finally the diagonal elements of the inverse Fisher matrix $F^{-1}_{ii}$ give the marginalized variance of the parameters. Therefore, the parameter uncertainty is given by
\begin{equation}
    \sigma_i = \sqrt{F^{-1}_{ii}}\,.
\end{equation}

We choose the fiducial model as the bestfit cosmology from Ref.\,\cite{Das:2020xke} for the SI mode in \three{} case with Planck 2018 TTTEEE+lowE data. Here we note that the SI mode posterior in SINU cosmology is not a perfect Gaussian. It is contiguous with the MI mode at smaller $\geff$. However, for the purpose of forecasting the upper limit, we expect the Fisher formalism to be applicable as long as we compute the derivatives within a small enough region around the fiducial point where the posterior can be taken to be approximately Gaussian. This is a reasonable assumption given the shapes of the SI mode in Figs.\,\ref{fig:tri-act} and \ref{fig:tri-act-planck}. We ignore the left side of the $\lgeff$ posterior. While computing the derivatives in Eq.(\ref{eq:fisher}), we checked the convergence with decreasing value of the step size. We found taking a step size of $1\%$ of the fiducial value is sufficient for all parameters. We use $\lmax=4000$ in all forecasts.

First, we investigate the constraining power of different multipoles. The diagonal elements of the terms on the right-hand side of Eq.(\ref{eq:fisher}) contain this information. In Fig.\,\ref{fig:fisher_info_geff}, we plot the diagonal term corresponding to $\lgeff$ as a function of $\ell$ for CMB-S4, CORE, and cosmic variance limited experiments. Looking at the last case, when the instrumental noise is negligible, most of the constraining power comes from the high-$\ell$ tail in the range $2000\lesssim\ell\lesssim4000$ with a substantial part coming from the intermediate multipoles $\ell\lesssim2000$. However, when instrumental noise is taken into account, the contribution from the high-$\ell$ region goes down significantly as $\mathbf{N}_\ell$ dominates at large $\ell$. Due to its better angular resolution, CMB-S4 can still access the multipoles above 2000 up to $\ell\lesssim2800$. In the case of CORE, the range is limited to $\ell\lesssim1900$.

When we look at the data from the current experiments (cf. Fig.\,\ref{fig:residuals_tt} and \ref{fig:residuals_ee}), the decisive multipole range for SINU is $500\lesssim\ell\lesssim1200$. It was also pointed out in Ref.\cite{Friedland:2007vv}. Such is the case because the current experiments do not have enough precision at small angular scales. However, this should be contrasted with the forecast for future CMB experiments discussed here which suggests that more information from the CMB spectra can be acquired from higher $\ell\gtrsim2000$ multipoles if small enough instrumental noise is achieved.

Fig.\,\ref{fig:fisher_corner} shows the 1$\sigma$ error ellipses for a few  parameters. From the panel of $\lgeff$, we see that CMB-S4 is expected to yield about three times better $1\sigma$ (68\% confidence level) upper limit compared to the latest and the final data from the Planck experiment. We quote the future sensitivities of $\lgeff$ in Table\,\ref{tab:fom}. We note that our results for CMB-S4 matches with the previous work\,\cite{Lancaster:2017ksf}. The $1\sigma$ errorbar for CORE is understandably bigger because of its lesser sensitivity. Finally, a cosmic variance-limited experiment with negligible instrumental noise would perform even better as expected. The effectiveness of the future experiments can also be measured by defining a meaningful Figure-of-merit (FoM). It has been defined in multiple ways in the literature. However, we follow the simple definition of FoM as the inverse of the volume of the 7-dimensional hyperellipsoid defined by the inverse Fisher matrix, FoM $= \sqrt{\mathrm{det}(F)}$\,\cite{Albrecht:2006um,Bassett:2009uv}. These FoMs are given in Table\,\ref{tab:fom}.
\begin{table}[h]
\begin{center}
\begin{tabular}{|l|P{3.5cm}|P{2.77cm}|P{2.77cm}|}
\hline
Experiment & Sensitivity for $\log_{10}[\geff/\mathrm{MeV^{-2}}]$ & Figure-of-merit (FoM) & FoM relative to CV-limited\\
\hline\hline
CMB-S4 & $<-1.87$ & $1.1\times10^{28}$ & $1.8\times10^{-3}$\\
\hline
CORE & $<-1.84$ & $3.9\times10^{26}$ & $6.4\times10^{-5}$\\
\hline
CV-limited & $<-1.9$ & $6.1\times10^{30}$ & 1\\
\hline
\end{tabular}
\end{center}
\caption{The future sensitivity of $\lgeff$ and overall figure-of-merit of the upcoming CMB experiments, and comparison with a cosmic variance-limited experiment. Note that the sensitivities are based on the SI mode bestfit point of Planck 2018 TTTEEE+lowE data.}\label{tab:fom}
\end{table}


\section{Discussion \& Outlook}
\label{sec:conclusion}
We revisit the study of the cosmology of self-interacting neutrinos using the new CMB data from the Atacama Cosmology Telescope (ACT) experiment in combination with the latest 2018 data from the Planck experiment. Prior cosmological data was known to yield two distinct modes in the neutrino interaction strength ($\geff$) -- the moderately interacting (MI) mode with a weaker coupling that merges with the \lcdm{} cosmology, and a strongly interacting (SI) mode, in which case, neutrinos decouples just prior to the matter-radiation equality epoch. However, the 2018 Planck polarization data strongly disfavored the SI mode\,\cite{Das:2020xke}. 

In this work, on the other hand, we find that the new ACT data favors SI over the MI mode irrespective of the number of coupled neutrino species. The reason behind this can be assigned to a certain feature in the ACT polarization data in the $500\lesssim\ell\lesssim1200$ multipole range that is not present in the Planck 2018 data. This conclusion corroborates with the findings of another work in Ref.\,\cite{Kreisch:2022zxp} in the flavor-universal case. Such a mismatch between the two datasets could be resolved with upcoming experiments that will measure the polarization data with more precision. We do not find a significant change in the value of the Hubble parameter. However, the combined ACT and Planck data prefer the SI mode over \lcdm{} for flavor-specific interaction.

We also studied the effect of adding neutrino mass to this analysis. We found slightly smaller values of the Hubble parameter $H_0$ and the dark energy density $\Omega_\Lambda$ as massive neutrinos change the total matter density of the Universe for all experiments. These changes are greater for Planck due to its large-scale data. There is no other qualitative difference between the massless and massive neutrino results proving the robustness of our baseline massless neutrino model discussed in Sec.\,\ref{sec:baseline}. 

Adding more free streaming radiation in the early Universe has the opposite effect of neutrino self-coupling on the CMB power spectra. It increases bestfit values of $H_0$ and $n_s$ but does not change the position of the SI mode in $\lgeff$. We note that none of these extended models yield a significantly large value of $H_0$ or change the position of the SI mode in $\geff$. On the other hand, when we vary the $\dneff$ for the coupled neutrinos, we find that $H_0$ and $n_s$ get slightly smaller relative to the fixed $\neff$ case. These changes are explained by the net decrease in $N_{\rm eff}$.

Finally, we did a Fisher analysis of the future prospect of a few upcoming/planned CMB experiments. We chose CMB-S4 and CORE as two benchmark terrestrial and space-based experiments, and use their publicly available data and technical design parameters. The arcminute level angular resolution of CMB-S4 will significantly improve the sensitivity at smaller angular scales. It will help constrain or detect signatures of neutrino self-interaction in the large multipole ($\ell\gtrsim2000$) tail of the CMB spectrum. The forecast sensitivity of CMB-S4 is roughly twice as good as CORE for $\lmax=4000$. This is due to the relatively lesser sensitivity of CORE at the small scale. However, we note that although CMB-S4 is an approved experiment and it is currently in the design phase, CORE is yet to be approved and its instrument requirements are subject to change in the future. Nonetheless, the complementarity of the science scopes of ground and space-based future experiments is expected to be crucial and will be key to continuing the legacy of CMB observation in the coming decades\,\cite{Delabrouille-Future-CMB}.

\section*{Acknowledgment}
The work of A.D. was supported by the U.S. Department of Energy under contract number DE-AC02-76SF00515 and Grant Korea NRF-2019R1C1C1010050. S.G. is supported by the U.S. National Science Foundation grant PHY2112540. The computational part of the work is supported by the Fermi National Accelerator Laboratory, managed and operated by Fermi Research Alliance, LLC under Contract No. DE-AC02-07CH11359 with the U.S. Department of Energy, Office of Science, Office of High Energy Physics. A.D. gratefully acknowledges the hospitality at
APCTP during the completion of this work. 
\begin{table}[t]
\begin{center}
\begin{tabular}[b]{l c}
\hline
\textbf{Numerical computation} & \\
\hline
Total core hour [$\mathrm{h}$]& 359,201\\
Thermal Design Power per core [$\mathrm{W}$]\,\cite{Intel}& 11.88\\
Total energy consumption for computations [$\mathrm{kWh}$] & 4266\\
Average emission of CO$_2$ in Illinois, US [$\mathrm{kg/kWh}$]\,\cite{IL}& 0.3\\
Total CO$_2$-emission for numerical computations [$\mathrm{kg}$] & 1280\\
Were the emissions offset? & \textbf{No}\\
\hline
\textbf{Travel} & \\
\hline
Total CO$_2$-emission for travel [$\mathrm{kg}$] & 0\\
\hline\hline
Total CO$_2$-emission [$\mathrm{kg}$] & 1280\\
\hline
\hline
\end{tabular}
  \caption{Approximate estimate of the total CO$_2$ emission during the course of this work.}\label{tab:co2}
\end{center}
\end{table}

\section*{Carbon footprint estimate}
As this work involved intensive numerical computation, in Table\,\ref{tab:co2}, we give an estimate of the breakdown and total CO$_2$ emission from the activities related to it\,\cite{emission}. The total emission is approximately 1280 kg, equivalent to about 5 hours of air travel. 

\appendix

\section{Temperature scaling of $\geff$}
\label{app:gefft}
\begin{figure}[t]
    \centering
    \includegraphics[width=\linewidth]{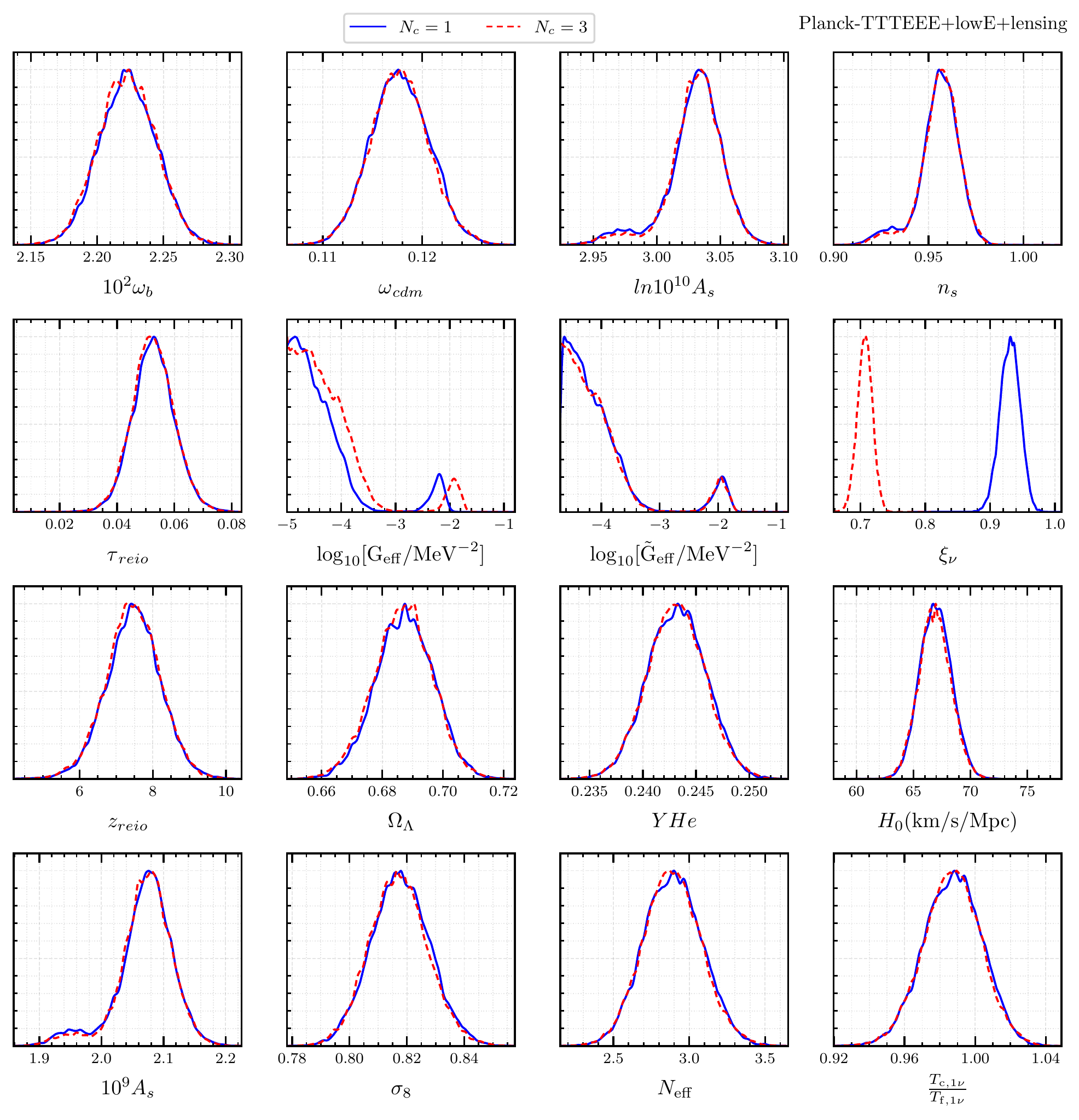}
    \caption{1D plots of cosmological parameters with self-interacting neutrino species with $N_c = 3 $ and $N_c = 1$ respectively where the temperature is allowed to vary. The constraints on $\lgefft$ and other \lcdm{} cosmological parameters are identical in both cases.}
    \label{fig:temp_comp_planck}
\end{figure}
In this section, we perform mcmc analyses of \three{} scenarios with varying temperatures for the Planck dataset. We choose two different methods which outline the dependence of $\lgeff$ on temperature. In the first case, we choose the number of coupled flavors $N_c = 3$, and in the latter scenario, we choose $N_c =1$. Fig.\,\ref{fig:temp_comp_planck} shows the 1D posterior distributions in these cases. The $N_c =1$ expectedly prefers a higher neutrino temperature $\xi_\nu$. As expected the constraint on $\lgeff$ is weaker in this scenario which follows from the arguments in sec.\,\ref{sec:results}. However, the scaled coupling constant $\lgefft$, as defined in Eq.(\ref{eq:pseudogeff}), is identical in both cases. Moreover, the other \lcdm{} cosmological parameter values are also identical in both cases. This shows that varying the temperature of neutrinos to account for varying $\neff$ produces identical constraints on \lcdm{} parameters as varying $\neff$ directly. However, the constraints on $\lgeff$ scale according to the temperature. We also show constraints on a derived parameter ${T_{{\rm c},1\nu}} / {T_{{\rm f},1\nu}}$ which is the ratio of coupled vs free streaming \emph{one-neutrino} temperature. $T_{{\rm c},1\nu}$ is defined as $T_{{\rm c},1\nu} \equiv T_{{\rm c}}/N_c^{1/4}$ and ${T_{{\rm f},1\nu}}$ is the free streaming neutrino temperature in SM.

\section{Parameter value for massive SINU}
\label{sec:mass_table}
\begin{table}[t]

    \centering
    \resizebox{\textwidth}{!}{
    \begin{tabular}{|c|P{2.6cm}P{2.6cm}|P{2.6cm}P{2.6cm}|P{2.6cm}P{2.6cm}|}
\hline
$\mathbf{2c+1f}$ &\multicolumn{2}{c}{ACTDR4+tauprior}\vline &\multicolumn{2}{c}{Planck-TTTEEE+lowE+lensing}\vline &\multicolumn{2}{c}{\makecell{Planck-TTTEEE+lowE+lensing\\+ACTDR4}}\vline\\
\hline
Parameters  & Massless $\nu$ & Massive $\nu$ & Massless $\nu$ & Massive $\nu$ & Massless $\nu$ & Massive $\nu$\\
\hline
$10^2 \omega_{\rm b}$ & $ 2.154\pm 0.031$ & $ 2.153\pm 0.030$ & $ 2.237^{+0.014}_{-0.016}$ & $ 2.234\pm 0.015$ & $ 2.236\pm 0.013$ & $ 2.234\pm 0.013$\\
$\omega_{\rm cdm}$ & $ 0.1202\pm 0.0038$ & $ 0.1206\pm 0.0039$ & $ 0.1200\pm 0.0012$ & $ 0.1204\pm 0.0012$ & $ 0.1203\pm 0.0012$ & $ 0.1207\pm 0.0012$\\
$100\theta{}_{s }$ & $ 1.0462^{+0.0018}_{-0.00039}$ & $ 1.0461^{+0.0019}_{-0.00039}$ & $ 1.042230^{+0.000020}_{-0.00070}$ & $ 1.042197^{+0.000036}_{-0.00068}$ & $ 1.0440^{+0.0020}_{-0.0022}$ & $ 1.0441^{+0.0017}_{-0.0021}$\\
$\ln10^{10}A_s$ & $ 3.007^{+0.025}_{-0.032}$ & $ 3.009^{+0.025}_{-0.033}$ & $ 3.037^{+0.020}_{-0.013}$ & $ 3.040^{+0.020}_{-0.013}$ & $ 3.025\pm 0.025$ & $ 3.026\pm 0.025$\\
$n_{s }$ & $ 0.979\pm 0.018$ & $ 0.978\pm 0.018$ & $ 0.9611^{+0.0078}_{-0.0034}$ & $ 0.9602^{+0.0077}_{-0.0035}$ & $ 0.955\pm 0.010$ & $ 0.953^{+0.015}_{-0.013}$\\
$\tau_{\rm reio}$ & $ 0.0595\pm 0.0098$ & $ 0.0597\pm 0.0097$ & $ 0.0536\pm 0.0073$ & $ 0.0544\pm 0.0073$ & $ 0.0546\pm 0.0058$ & $ 0.0551\pm 0.0059$\\
$\log_{10} [{\rm G_{eff}} /  {\rm MeV}^{-2}]$ & $ -2.04^{+0.66}_{+0.22}$ & $ -2.07^{+0.69}_{+0.23}$ & $< -3.82$ & $< -3.84$ & $ -2.7\pm 1.2$ & $ -2.6^{+1.1}_{-1.3}$\\
\hline
$H_0 ({\rm km/s/Mpc})$ & $ 68.5\pm 1.5$ & $ 67.8\pm 1.5$ & $ 68.02^{+0.55}_{-0.65}$ & $ 67.33^{+0.52}_{-0.69}$ & $ 68.50\pm 0.64$ & $ 67.83\pm 0.64$\\
$\sigma_8$ & $ 0.838\pm 0.016$ & $ 0.826\pm 0.016$ & $ 0.8240^{+0.0057}_{-0.0067}$ & $ 0.8129\pm 0.0062$ & $ 0.8290\pm 0.0059$ & $ 0.8179\pm 0.0058$\\
\hline
$\chi^2 - \chi^2_{\Lambda{\rm CDM}}$ & $-6.92$ & $-6.99$ & $0.17$ & $1.26$ & $-5.31$ & $-4.9$\\
\hline
$\Delta{\rm AIC}$ & $-4.92$ & $-4.99$ & $2.17$ & $3.26$ & $-3.31$ & $-2.9$\\
\hline
\end{tabular}
    }
    \newline
    \vspace*{2ex}
    \newline
    \resizebox{\textwidth}{!}{
    \begin{tabular}{|c|P{2.6cm}P{2.6cm}|P{2.6cm}P{2.6cm}|P{2.6cm}P{2.6cm}|}
\hline
$\mathbf{1c+2f}$ &\multicolumn{2}{c}{ACTDR4+tauprior}\vline &\multicolumn{2}{c}{Planck-TTTEEE+lowE+lensing}\vline &\multicolumn{2}{c}{\makecell{Planck-TTTEEE+lowE+lensing\\+ACTDR4}}\vline\\
\hline
Parameters  & Massless $\nu$ & Massive $\nu$ & Massless $\nu$ & Massive $\nu$ & Massless $\nu$ & Massive $\nu$\\
\hline
$10^2 \omega_{\rm b}$ & $ 2.155\pm 0.030$ & $ 2.154\pm 0.030$ & $ 2.237\pm 0.014$ & $ 2.234\pm 0.015$ & $ 2.235\pm 0.013$ & $ 2.233\pm 0.013$\\
$\omega_{\rm cdm}$ & $ 0.1197\pm 0.0037$ & $ 0.1200\pm 0.0037$ & $ 0.1200\pm 0.0012$ & $ 0.1204\pm 0.0012$ & $ 0.1200\pm 0.0011$ & $ 0.1205\pm 0.0011$\\
$100\theta{}_{s }$ & $ 1.0443\pm 0.0011$ & $ 1.0443\pm 0.0011$ & $ 1.0423^{+0.0012}_{-0.00091}$ & $ 1.04228^{+0.00015}_{-0.00091}$ & $ 1.0432^{+0.0010}_{-0.0013}$ & $ 1.0432^{+0.0010}_{-0.0013}$\\
$\ln10^{10}A_s$ & $ 3.031\pm 0.025$ & $ 3.032^{+0.023}_{-0.026}$ & $ 3.036^{+0.017}_{-0.016}$ & $ 3.039^{+0.017}_{-0.015}$ & $ 3.036^{+0.014}_{-0.019}$ & $ 3.038^{+0.013}_{-0.019}$\\
$n_{s }$ & $ 0.994^{+0.017}_{-0.013}$ & $ 0.993^{+0.016}_{-0.013}$ & $ 0.9610^{+0.0065}_{-0.0051}$ & $ 0.9602^{+0.0066}_{-0.0048}$ & $ 0.9601^{+0.0052}_{-0.0072}$ & $ 0.9590^{+0.0056}_{-0.0072}$\\
$\tau_{\rm reio}$ & $ 0.0598\pm 0.0097$ & $ 0.0601\pm 0.0096$ & $ 0.0536\pm 0.0073$ & $ 0.0545\pm 0.0073$ & $ 0.0551\pm 0.0058$ & $ 0.0555\pm 0.0058$\\
$\log_{10} [{\rm G_{eff}} / {\rm MeV}^{-2}]$ & $ -2.5\pm 1.1$ & $ -2.4\pm 1.1$ & $< -2.84$ & $< -2.92$ & $ -2.4^{+1.1}_{-1.3}$ & $ -2.4^{+1.1}_{-1.3}$\\
\hline
$H_0 ({\rm km/s/Mpc})$ & $ 68.1\pm 1.5$ & $ 67.5\pm 1.5$ & $ 68.04\pm 0.58$ & $ 67.35\pm 0.57$ & $ 68.32^{+0.55}_{-0.49}$ & $ 67.63\pm 0.52$\\
$\sigma_8$ & $ 0.838\pm 0.015$ & $ 0.826^{+0.014}_{-0.016}$ & $ 0.8235\pm 0.0062$ & $ 0.8124\pm 0.0061$ & $ 0.8275\pm 0.0059$ & $ 0.8162\pm 0.0057$\\
\hline
$\chi^2 - \chi^2_{\Lambda{\rm CDM}}$ & $-4.08$ & $-4.04$ & $0.13$ & $1.22$ & $-5.16$ & $-4.79$\\
\hline
$\Delta{\rm AIC}$ & $-2.08$ & $-2.04$ & $2.13$ & $3.22$ & $-3.16$ & $-2.79$\\
\hline
\end{tabular}
    }
        \caption{Mean values and 68\% confidence limits for massless-$\nu$ and massive-$\nu$ cases for \two{} and \one{}. Note that the values and limits are for the full sample where we have not separated out the MI and SI modes. We also show the $\chi^2$ difference from the bestfit 6-parameter $\Lambda$CDM model and the corresponding $\Delta$AIC}\label{tab:act_massive}
\end{table}
\section{Triangle plots for ACT and Planck+ACT analysis}
\label{app:triplots}
\begin{figure}
    \centering
    \includegraphics[width = \linewidth]{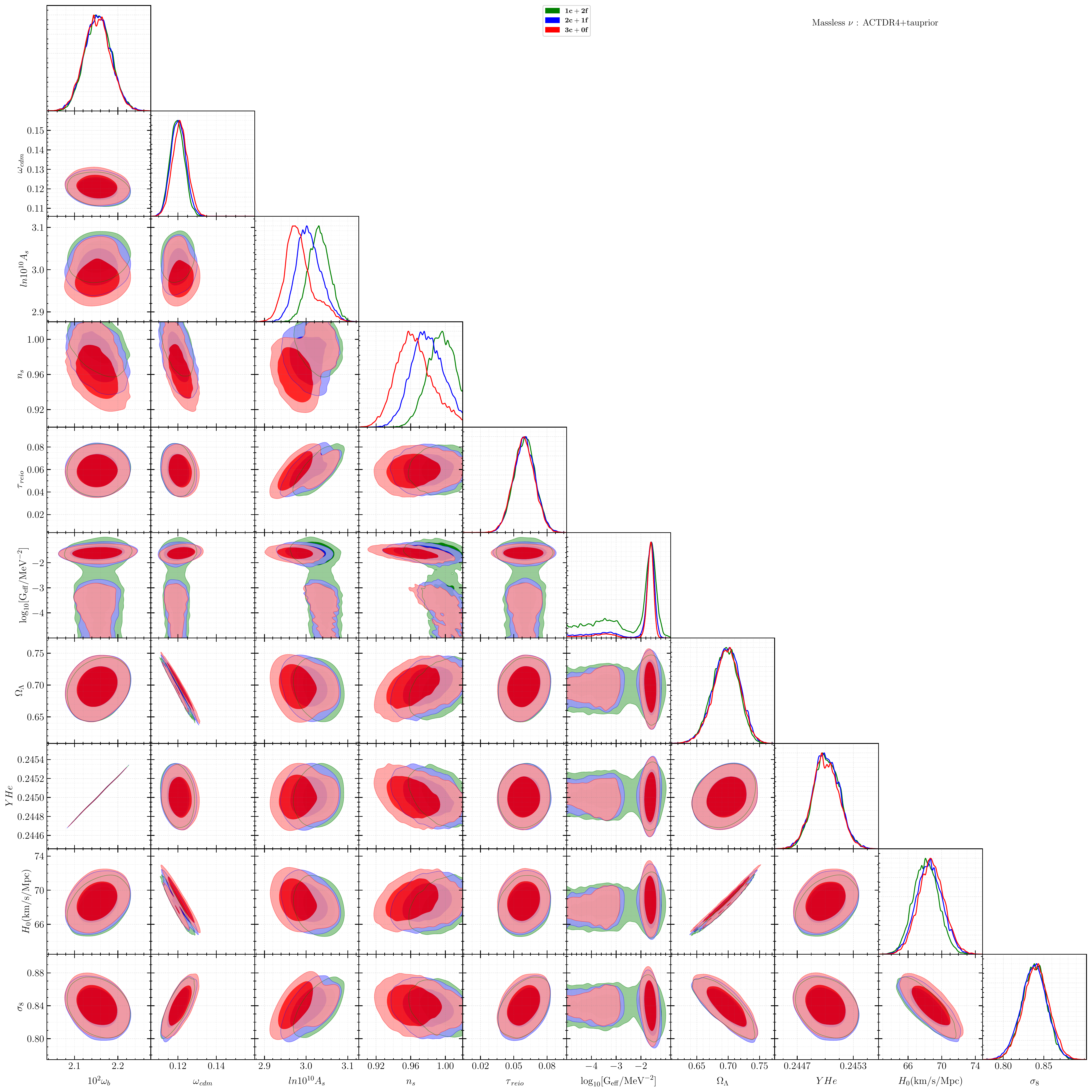}
    \caption{Triangle plot for an extended set of cosmological parameters for the baseline model analysis with ACT dataset.}
    \label{fig:triall_baseline_ACT}
\end{figure}
\begin{figure}
    \centering
    \includegraphics[width = \linewidth]{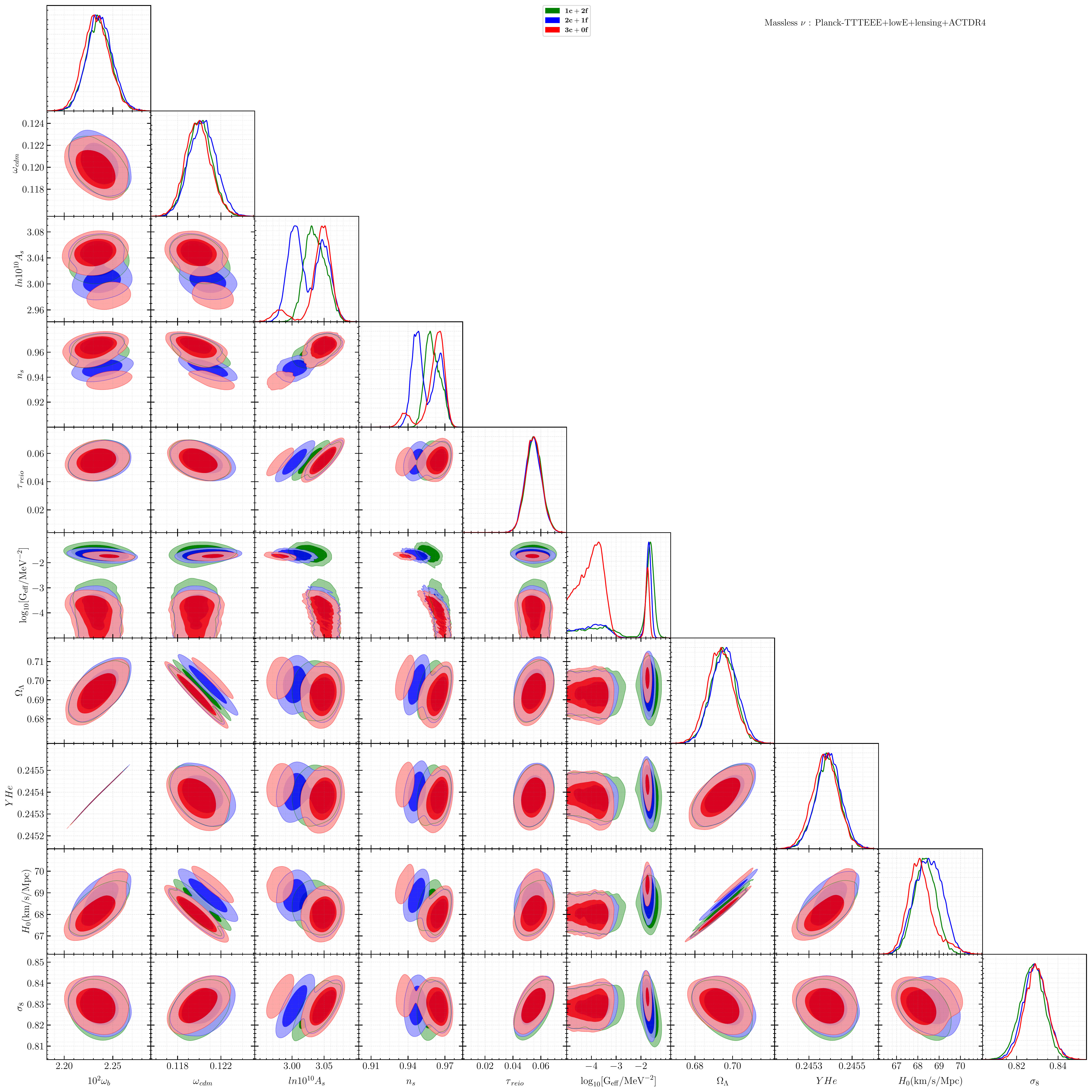}
    \caption{Triangle plot for an extended set of cosmological parameters for the baseline model analysis with Planck + ACT dataset.}
    \label{fig:triall_baseline}
\end{figure}
\begin{figure}
    \centering
    \includegraphics[width = \linewidth]{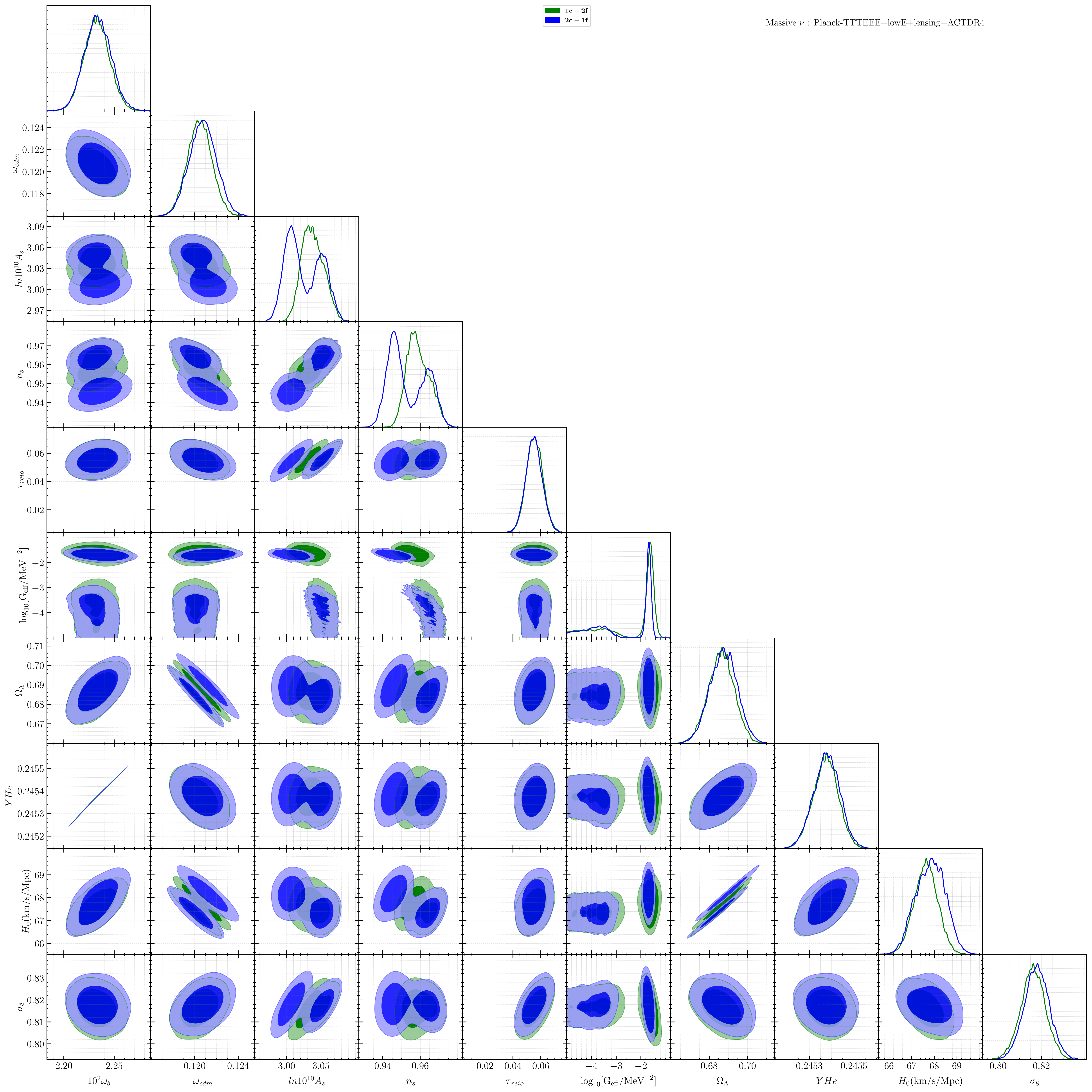}
    \caption{Triangle plot for an extended set of cosmological parameters for the analysis with massive neutrino for Planck + ACT dataset.}
    \label{fig:triall_massive}
\end{figure}
\begin{figure}
    \centering
    \includegraphics[width = \linewidth]{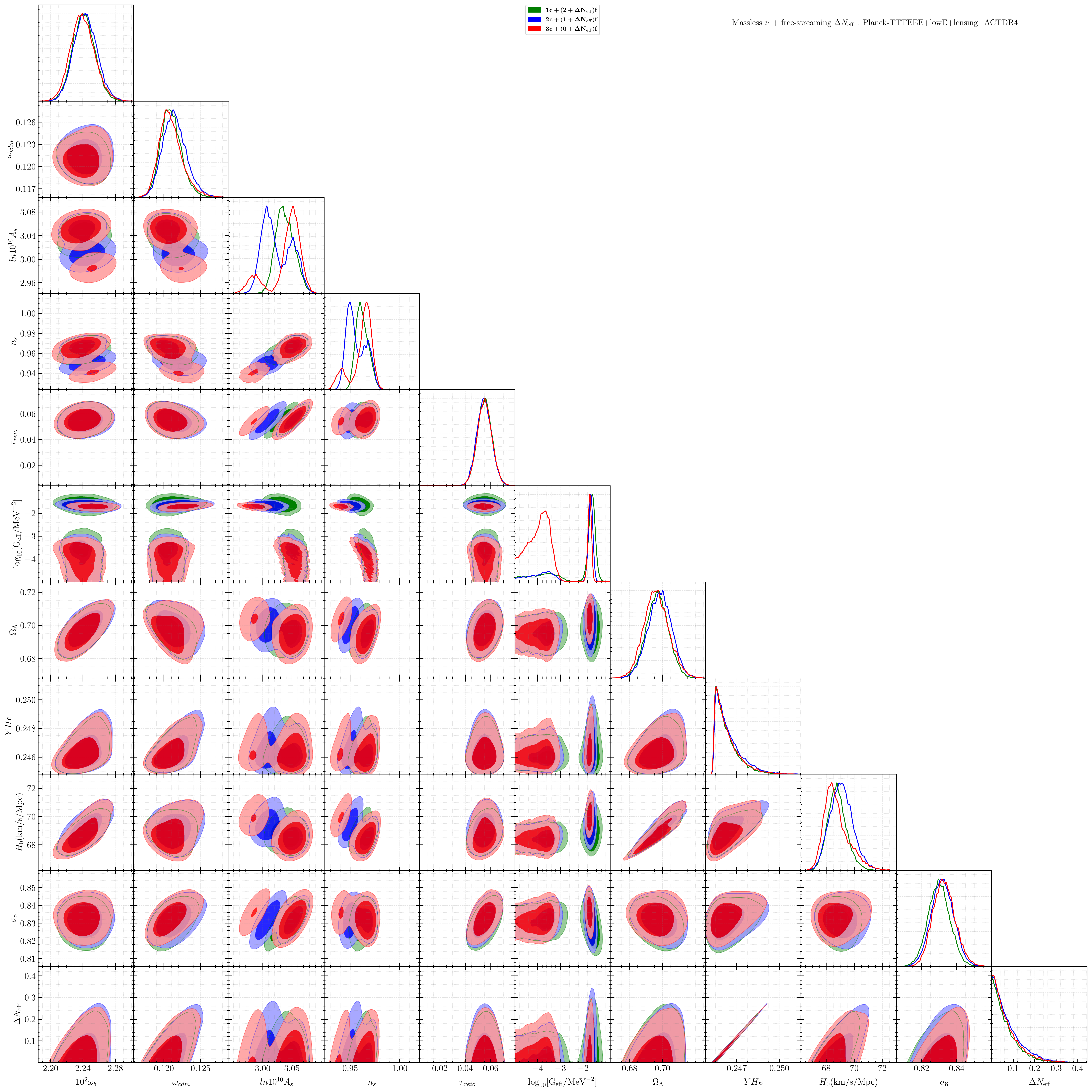}
    \caption{Triangle plot for an extended set of cosmological parameters for additional free-streaming $\Delta N_{\rm eff}$ analysis with Planck + ACT dataset.}
    \label{fig:triall_fsrad}
\end{figure}
\begin{figure}
    \centering
    \includegraphics[width = \linewidth]{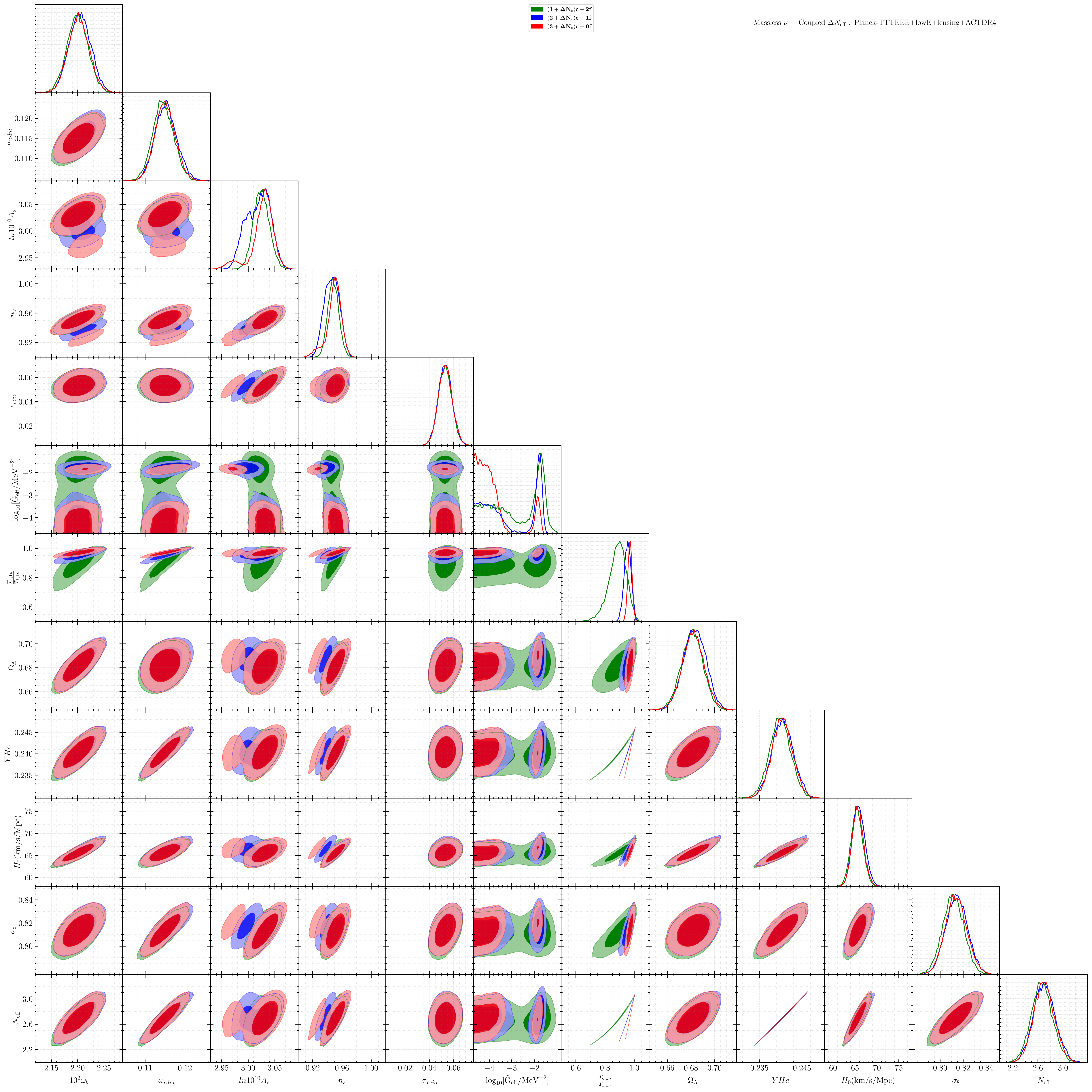}
    \caption{Triangle plot for an extended set of cosmological parameters for additional coupled $\Delta N_{\rm eff}$ analysis with Planck + ACT dataset.}
    \label{fig:triall_nfsrad}
\end{figure}

\bibliographystyle{jhep}
\bibliography{sinu}

\end{document}